\documentclass[twocolumn]{aastex701}
\usepackage{amsmath,amssymb,bm,graphicx,multirow,nccmath}
\usepackage[T1]{fontenc}

\newcommand{\Fischer}{F25}
\newcommand{\dd}{{\rm d}}
\newcommand{\DF}{\textsc{df}}

\renewcommand{\vec}{\bm}
\newcommand{\uvec}[1]{\hat{\bm{#1}}}

\newcommand{\Sec}{Section~}
\newcommand{\Appx}{Appendix~}
\newcommand{\Fig}{Figure~}
\newcommand{\Eqn}{Equation~}
\newcommand{\Tab}{Table~}
\renewcommand{\eqref}[1]{\Eqn\ref{#1}}

\newcommand{\Secs}{Sections~}

\newcommand{\Figs}{Figures~}
\newcommand{\Eqns}{Equations~}

\begin{document}

\title{A New Method to Simulate Dark Matter--Baryon Interactions\\and Application to an Isolated Disk Galaxy}

\author{Connor Hainje}
\affiliation{Center for Cosmology and Particle Physics, Department of Physics, New York University}
\email{connor.hainje@nyu.edu}

\author{Glennys R.~Farrar}
\affiliation{Center for Cosmology and Particle Physics, Department of Physics, New York University}
\email{gf25@nyu.edu}

\begin{abstract}
We report on a new method for incorporating interactions between dark matter (DM) and baryons in cosmological simulations, capable of handling the challenging regime in which the dark matter particle mass is comparable to or lighter than the baryon mass.
The method hybridizes two distinct approaches: gas particles receive momentum and energy transfer according to a mean-field calculation while DM particles undergo Monte Carlo scatterings.
These approaches are derived from the Boltzmann equation and shown to be statistically equivalent.
We present an open-source implementation of this method in the simulation code GIZMO.
As a first application, we investigate the effects of DM--baryon interactions on an isolated Milky Way-like disk galaxy for dark matter having twice the proton mass, which roughly maximizes the average energy transfer per collision.
For cross sections of order 1 barn ($10^{-24} \ {\rm cm}^2$), these interactions cause strong changes to the mass distribution in the center of the galaxy in less than 1 Gyr, even when bar formation is suppressed by hand.
For cross sections typical of hadronic interactions ($\lesssim 30 \ {\rm mb}$), high-fidelity galaxy formation simulations will be needed to assess the effects on observable features of galaxies.
\end{abstract}


\section{Introduction}
\label{sec:intro}

In the standard Lambda Cold Dark Matter ($\Lambda$CDM) paradigm, dark matter (DM) interacts only via gravity.
This yields remarkably successful predictions on cosmological scales \citep[e.g.,][]{Planck2020}, but tensions have been reported on smaller scales \citep[e.g.,][]{BullockBoylan-Kolchin2017, Kaplinghat+2020}.
Furthermore, some particle physics models imply potentially observable impacts of non-gravitational dark matter interactions: either DM self-interactions \citep[SIDM; e.g.,][]{SpergelSteinhardt2000,TulinYu2018,Adhikari+2025} or DM interactions with the Standard Model \citep[e.g.,][]{Dvorkin+2014,Farrar2022}.
In this work, we focus on dark matter--baryon interactions, for which there has been a dearth of computational tools.

Dark matter--baryon (DM--b) interactions are a feature of many particle models of dark matter.
The most common such model is the Weakly Interacting Massive Particle (WIMP), although it is not very interesting observationally due to its heavy mass, small cross section with baryons, and the severe constraints imposed by accelerator limits and direct detection experiments \citep[e.g.,][]{Aprile+2023-XENONnT,Aalbers+2025-LZ}.
We focus on dark matter with more observationally impactful interactions with baryons, such as the regime of light mass DM with moderate-to-strong interactions with baryons that is challenging to directly detect.
In the GeV mass range, the very sensitive, deep underground direct detection experiments like SuperCDMS can only constrain small cross sections of $10^{-35}$ to $10^{-29} \ {\rm cm}^2$, because for, larger cross sections, scattering in the overburden depletes the DM energy to below the detection threshold~\citep{ZaharijasFarrar2005, MahdawiFarrar2017, CDMSlomass23}, leaving the range $\gtrsim 10^{-28} \ {\rm cm}^2$ nearly unconstrained \citep{xf23}.\footnote{The upscattering of DM by cosmic rays can enhance the sensitivity of direct detection experiments \citep{AlveyBringmann23}, which could significantly constrain the parameter space in this range; however, the applicability of the constraints is limited~\citep{Bell+CRDM24} and there can be loopholes when non-perturbative energy losses in propagation through the Earth are taken into account~\citep{xf23}.}

A complementary approach to constraining DM--b interactions is through their astrophysical impacts.
For example, \citet{wf21} placed limits by requiring that the heating or cooling due to DM--b interactions not exceed the radiative cooling rate of gas in gas-rich dwarf galaxies.
DM--b interactions can also affect the structure of galaxies \citep[e.g.,][]{Dvorkin+2014,Buen-Abad+2022}, primarily in two ways:
    by introducing viscosity before recombination and thereby suppressing the power spectrum on small scales,
    and by changing the formation and evolution of galaxies.
The pre-recombination imprints have been used to place limits on interaction parameters by comparing against measurements of
    the CMB \citep{GluscevicBoddy2018,BoddyGluscevic2018,Boddy+2018,Ali-Haimoud+2024},
    the Lyman-$\alpha$ forest \citep{Dvorkin+2014,Xu+2018},
    large-scale structure \citep{He+2023,He+2025},
    and the abundance of Milky Way satellites \citep{Nadler+2021,Maamari+2021,Buen-Abad+2022}.
For $m_\chi = 2 \, m_p$, these studies place upper bounds ranging
    from $\sigma \lesssim 10^{-25} \, {\rm cm}^2$ (CMB and BAO)
    to $\sigma \lesssim \text{few} \times 10^{-28} \, {\rm cm}^2$ (Lyman-$\alpha$ and Milky Way satellites)
    \citep[summarized in][]{Buen-Abad+2022}.

However, there are two important caveats to these galactic structure analyses:
    they assume that the dark matter is an ideal fluid, and
    they neglect the effects of interactions during the non-linear regime of structure formation.
Avoiding these assumptions requires resolving the interplay between DM--b interactions and baryonic physics in detail through a cosmological simulation, which is very challenging but could be very important.
For example, it is known that the departure of the dark matter from an ideal fluid can lead to errors as large as 160\% in the heating rate \citep{SeherGandhiAli-Haimoud2022}.
At late times, DM--b interactions can provide a new, diffuse source of heating, which may moderate star formation and affect feedback from active galactic nuclei and supernovae.
DM--b interactions could also change the shape and kinematics of dark matter halos.
These effects cannot be treated in isolation.
Further, $\Lambda$CDM predictions in this regime commonly rely on ``gastrophysics'' prescriptions like subgrid models, many of which are tuned to reproduce the observed distribution of galaxies, and so potentially can be re-tuned to reproduce the observations in the presence of DM--b interactions.
For example, \citet{vanDaalen+2020} finds that small-scale suppression of the matter power spectrum varies widely among galaxy formation models and can reach up to 30\%, while
\citet{CAMELS2021} finds that varying feedback parameters within a single model by a factor of a few can have a comparably large impact.
Since DM--b interactions may affect galaxies through the same channels as feedback (such as suppressing power on small scales), careful consideration of this re-tuning will be important when using observations to test for and limit DM--b interactions.

The purpose of our work is to develop tools to enable an accurate treatment of the effects of dark matter--baryon interactions during galaxy formation and evolution, without making use of the ideal fluid approximation.
We have developed a new method for incorporating DM--b interactions into hydrodynamic cosmological simulations, which we have implemented as a module in the code GIZMO \citep{Hopkins2015-GIZMO}.
We focus on elastic scatterings between non-relativistic dark matter and baryon particles using the standard phenomenological, isotropic power-law cross section $\sigma (v) = \sigma_0 \, (v/c)^n$, where $v \ll c$ is the relative velocity \citep[e.g.,][]{Dvorkin+2014,Ali-Haimoud2019,Buen-Abad+2022}, though we note that our method can be applied to general cross section dependencies on $v$ and scattering angle.
Because electrons constitute such a small fraction of the mass of ordinary matter, and the limits on dark matter interactions with electric charge are very stringent \citep[e.g.][]{wf21}, we neglect lepton couplings entirely, although they could readily be incorporated.  For simplicity of exposition in this paper, we set the DM--baryon cross section to be the same for all nuclei; a more detailed treatment would be straightforward to implement.

The method we have developed hybridizes two statistically equivalent but qualitatively different approaches:
the dark matter is treated in a Monte Carlo fashion, where simulation particles probabilistically undergo scatterings,
and the baryons are treated with a mean-field approach, where the statistical expectations for the transfer rates of bulk velocity and thermal energy are applied.
These are both derived directly from the Boltzmann equation and proven to conserve energy and momentum in expectation.

During the preparation of this article, we learned of a parallel effort to implement DM--b interactions in $N$-body simulations \citep{Fischer+2025} (\Fischer{} below).
The advantages and disadvantages of that approach relative to ours are discussed in Section~\ref{sec:comparison}.
The most significant differences are that the \Fischer{} approach conserves energy and momentum exactly, where in ours the conservation is only on average.
However, the \Fischer{} approach requires the dark matter and baryon macroparticle masses to satisfy $M_\chi/M_b < 3/28 \, (m_\chi/m_b) \approx 0.1 \, m_\chi/{\rm GeV}$ (Equation~12 in \Fischer{}),
so employing this method in the $m_\chi \lesssim$ few GeV regime requires a much larger number of dark matter macroparticles than for non-interacting DM.\footnote{%
For reference, state-of-the-art cosmological simulations adopt $M_\chi/M_b \approx 5$ \citep[e.g.,][]{Hopkins+2018-FIRE2,Nelson+2019-TNG,Nelson+2024-TNG-Cluster}, and some isolated galaxy sims choose ratios as high as $M_\chi/M_b \approx 100$ \citep{Kim+2016-AGORA-II}.}
We note that \Fischer{} provides an alternative treatment for frequent, small-angle scatterings (e.g., from highly forward cross sections) that avoids this macroparticle mass restriction.

Our method thus provides for the first time a practical tool for investigating dark matter--baryon interactions for a dark matter mass close to that of hydrogen and helium, where energy and momentum transfer in individual collisions is maximally efficient.
A concrete example of such a model is Sexaquark Dark Matter, in which the dark matter particle is a six-quark state ($uuddss$) whose binding is sufficiently deep that it is effectively stable; see \citet{Farrar2022} and references therein for a review of the constraints from particle physics, astrophysics, and cosmology, and \citet{MooreProfumoSDM25} for additional discussion.

As an initial application of our code, we simulate the isolated Milky Way-like disk galaxy of the AGORA project \citep{Kim+2016-AGORA-II} with a dark matter particle having twice the proton mass ($2 \, m_p$) and velocity-independent cross section.
We find that sufficiently strong dark matter--baryon interactions induce significant changes to the morphology of the center of the galaxy, driving gas and dark matter inflows toward the center.
For potentially realistic DM--b cross sections of order tens of millibarns and below, assessing the impact of dark matter--baryon interactions will require more detailed treatment of star formation, feedback, and the hierarchical formation of galaxies than we undertake in the present work.

The article is structured as follows.
\Sec\ref{sec:method} gives the relevant scattering theory and the algorithms used to implement our approach in code.
\Sec\ref{sec:tests} briefly discusses the tests we perform to validate our implementation.
\Sec\ref{sec:galaxy} presents the key results of our simulation of the AGORA disk galaxy.
\Sec\ref{sec:conclusions} concludes.
Appendices provide derivations and details about our test suite.

\section{Method}
\label{sec:method}

In this section, we derive our method and discuss our implementation.
We begin by establishing the Boltzmann equation of our simulations using three core assumptions that underpin $N$-body simulations.
We use this Boltzmann equation to derive both
    a probabilistic scattering treatment for dark matter, and
    the mean acceleration and heating rates for baryons.
Then, we describe our implementation of this method in the simulation code GIZMO, and finally we provide a detailed comparison of our method with that of \citet{Fischer+2025}.

We note that our method treats only interactions between dark matter and gas, not between dark matter and stars.
The primary reason for this is that the volume-filling fraction of stars is very small, so the scattering rate of dark matter particles with stars is many orders of magnitude smaller than that with gas particles.\footnote{%
    Assuming stars have a number density of $n_\star \sim 0.1 \ {\rm pc}^{-3}$ (in the solar neighborhood) and a geometric cross section of $\sigma_\star \sim \pi \, R_\odot^2$, and the ISM has a number density around $n_{\rm gas} \sim 1 \ {\rm cm}^{-3}$, the ratio of the gas scattering rate to the star scattering rate is
    $(n_{\rm gas} \, \sigma) / (n_\star \, \sigma_\star) \sim 10^9 \times (\sigma / 100 \ {\rm mb}) $.
}
There are some physical effects that are predicted to result from interactions between dark matter and stars \citep[dark matter capture or stellar heating; see, e.g.,][]{John+2024}.
However, these generally operate on long timescales or on objects that are not resolved in our simulation (individual stars, neutron stars), and their effects should be subdominant to galaxy-scale momentum and energy transfer, so we neglect them here.

\subsection{Boltzmann equation}

We consider elastic $2 \leftrightarrow 2$ collisions between non-relativistic dark matter particles (of mass $m_\chi$) and baryons (of mass $m_B$) with differential cross section $\dd \sigma / \dd \Omega (v_{\rm rel}, \theta)$.
In any individual scatter with initial velocities $\vec{v}_\chi, \vec{v}_B$, the outgoing velocities are given by
\begin{align}
    \vec{v}_\chi '
    &\equiv \vec{S}_\chi
    =
    \vec{v}_\chi +
    \mfrac{m_B}{m_\chi + m_B} \,
    v_{\chi B} \,
    \big[
        \uvec{v}_{\chi B}'
        - \uvec{v}_{\chi B}
    \big] , \label{eq:scatter_chi} \\
    \vec{v}_B '
    &\equiv \vec{S}_B
    =
    \vec{v}_B -
    \mfrac{m_\chi}{m_\chi + m_B} \,
    v_{\chi B} \,
    \big[
        \uvec{v}_{\chi B}'
        - \uvec{v}_{\chi B}
    \big] , \label{eq:scatter_B}
\end{align}
where $\vec{v}_{\chi B} \equiv \vec{v}_\chi - \vec{v}_B$,
$v_{\chi B} = |\vec{v}_{\chi B}|$,
$\uvec{v}_{\chi B} = \vec{v}_{\chi B} / |\vec{v}_{\chi B}|$,
and $\uvec{v}_{\chi B}'$ is a new direction.
The scattering angle $\theta$ describes the angle between the two unit vectors $\uvec{v}_{\chi B}$ and $\uvec{v}_{\chi B}'$: $\cos\theta = \uvec{v}_{\chi B} \cdot \uvec{v}_{\chi B}'$.
We introduce $\vec{S}_\chi$ and $\vec{S}_B$, which may be thought of as functions of $\vec{v}_\chi, \vec{v}_B, \Omega$, as a short notation representing this transformation.

In the continuum limit, dark matter and baryons are each described by a distribution function\footnote{%
    We use the convention $\int \dd^3 x \, \dd^3 v \, f(\vec{x}, \vec{v}) = \mathcal{N}$, the total number of particles.
    The number density is then given by $n(\vec{x}) = \int \dd^3 v \, f(\vec{x}, \vec{v})$.%
} (\DF) $f(\vec{x}, \vec{v})$ whose evolution is governed by the Boltzmann equation:
\begin{align}
    \frac{\dd f_\chi}{\dd t} (\vec{v}_\chi)
    &=
    \int \dd^3 v_B \, \dd \Omega \, 
    \frac{\dd \sigma}{\dd \Omega} \,
    v_{\chi B} \,
    \label{eq:boltzmann} \\ \nonumber
    &\qquad\times
    \big[
        f_\chi (\vec{S}_\chi) \, f_B (\vec{S}_B)
        - f_\chi (\vec{v}_\chi) \, f_B (\vec{v}_B)
    \big] ,
\end{align}
where the dependence on $\vec{x}$ is implicit.
The right-hand side of this equation is known as the collision integral.

Our goal is to implement the effects of these collisions in a Lagrangian-picture hydrodynamic cosmological simulation.
Such simulations are built on three approximations, which we will use to derive our method.

The first approximation made by all $N$-body simulations is that the dark matter is a ``softened $N$-body component'';
namely, a number of Monte Carlo samples are drawn from some true initial \DF{} $f_\chi$ and spatially convolved with a smoothing kernel $W(r, h)$,\footnote{%
    We assume throughout that any valid smoothing kernel $W(r, h)$ integrates to 1 and is zero (or else negligible) for $r > h$.
} producing the following approximate \DF{}
\begin{equation}
    \hat{f}_\chi (\vec{x}, \vec{v})
    =
    \sum_{i=1}^{N_\chi} \frac{M_i}{m_\chi} \,
    W (|\vec{x} - \vec{x}_i|, h_i) \,
    \delta (\vec{v} - \vec{v}_i) ,
    \label{eq:macroparticle-DF_DM}
\end{equation}
where $\delta(\vec{v})$ is the three-dimensional Dirac delta in velocity space.
This can be understood as representing the dark matter by $N_\chi$ ``macroparticles'', each having a
mass $M_i$,
position $\vec{x}_i$,
smoothing length $h_i$, and
velocity $\vec{v}_i$.

The second approximation is that the gas is represented by a thermalized softened $N$-body component.
By this, we mean that the gas is decomposed into macroparticles like dark matter, but each macroparticle has a bulk velocity and internal energy describing the velocity distribution, which is locally Maxwell-Boltzmann at all times.
Therefore, the gas is represented by the approximate \DF{}\footnote{%
    We note that this form of the baryon \DF{} is valid for any hydrodynamic solver.
    By computing the moments of the baryon--baryon Boltzmann equation for this \DF{}, one can derive the fluid equations for standard smoothed particle hydrodynamics \citep[e.g.,][]{Monaghan1992}.
    More modern solvers, such as the moving Voronoi mesh of AREPO \citep{Springel2010} or the meshless methods of GIZMO \citep{Hopkins2015-GIZMO}, generally assume more sophisticated geometries for the discretization of the gas.
    However, in these simulations the resulting particle distribution is then taken to have the form of \Eqn\ref{eq:macroparticle-DF_B} for other applications, such as solving gravity, and so we take this form for solving interactions with dark matter.}
\begin{equation}
    \hat{f}_B (\vec{x}, \vec{v})
    =
    \sum_{j=1}^{N_B} \frac{M_j}{\mu_j} \,
    W (|\vec{x} - \vec{x}_j|, h_j) \,
    \mathcal{G} (\vec{v} - \vec{V}_j, T_j / \mu_j) ,
    \label{eq:macroparticle-DF_B}
\end{equation}
where $\mathcal{G}$ is the probability density function of a three-dimensional isotropic Gaussian---i.e., the Maxwell-Boltzmann distribution.
In simulation, the gas is represented by $N_B$ macroparticles, each having a
mass $M_j$,
position $\vec{x}_j$,
smoothing length $h_j$,
bulk velocity $\vec{V}_j$, and
temperature $T_j$.
Gas macroparticles can also consist of different species of particles; we approximate the dark matter as having equal cross section with all species, so all that matters is the mean molecular weight, $\mu_j$.

The macroparticle \DF{}s for dark matter and gas have a similar form.
For brevity, we introduce the following notation:
\begin{equation}
    \hat{f} (\vec{x}, \vec{v}) =
    \sum_i n_i (\vec{x}) \, \nu_i (\vec{v}),
\end{equation}
where $n_i$ is a number density and $\nu_i$ is a probability density function for the velocities, equal to either a Dirac delta (for dark matter) or a Maxwell-Boltzmann distribution (for gas).

The third approximation underpinning these simulations is made by assuming that the time-evolution of the true \DF{}s $f_\chi, f_B$ is well-approximated even when enforcing that the form of the macroparticle \DF{}s $\hat{f}_\chi, \hat{f}_B$ is preserved.
In other words, after a timestep, the forms of \Eqns\ref{eq:macroparticle-DF_DM} and~\ref{eq:macroparticle-DF_B} can be imposed on the dark matter and gas, albeit with updated values of the parameters.
We make this point explicit because the full evolution of the \DF{} according to the Boltzmann equation would distort the functional form of the discretization---for example, the softened macroparticles would experience tidal forces and become stretched and sheared---but this is regularly ignored.
For our purposes, this assumption means that evolution according to the collision integral only updates the local velocity distributions of the individual particles.
Thus, the Boltzmann equation is
\begin{align}
    &\sum_i
    n_i (\vec{x}) \,
    \frac{\dd \nu_i}{\dd t} \bigg|_{\vec{v}_\chi}
    \\ \nonumber
    &\qquad =
    \sum_{i,j} n_i (\vec{x}) \, n_j (\vec{x}) \,
    \int \dd^3 v_B \, \dd \Omega \,
    \frac{\dd\sigma}{\dd\Omega} \, v_{\chi B} \,
    \\ \nonumber &\qquad\qquad\times
    \big[
    \nu_i (\vec{S}_\chi) \, \nu_j (\vec{S}_B)
    -
    \nu_i (\vec{v}_\chi) \, \nu_j (\vec{v}_B)
    \big] ,
\end{align}
and similarly for the baryons.

The form of this equation suggests a matching procedure: term $i$ on the left-hand side can be set equal to term $i$ on the right-hand side.
By doing this matching, we are making an additional assumption: that the phase space densities of macroparticles of the same type do not mix, so each macroparticle evolves as if it is the only particle of its type at its location.
This assumption is reasonable as long as no particles have identical positions $\vec{x}$, which is already required for the validity of the $N$-body approach.
Hence, individual macroparticles evolve according to independent collision integrals:
\begin{align}
    \frac{\dd f_{\chi}^i}{\dd t} (\vec{v}_\chi)
    &=
    \int \dd^3 v_B \, \dd \Omega \, 
    \frac{\dd \sigma}{\dd \Omega} \,
    v_{\chi B} \,
    \label{eq:boltzmann-individual} \\ \nonumber
    &\qquad\times
    \big[
        f_{\chi}^i (\vec{S}_\chi) \, \hat{f}_B (\vec{S}_B)
        - f_{\chi}^i (\vec{v}_\chi) \, \hat{f}_B (\vec{v}_B)
    \big] ,
\end{align}
with $\vec{x}$-dependence implicit and defining $f_\chi^i(\vec{x}, \vec{v}) \equiv n_i (\vec{x}) \, \nu_i (\vec{v})$.
This equation applies analogously to the baryons.

\subsection{Dark matter evolution}

In order to determine how to update the velocities of the dark matter particles, consider a finite timestep $\Delta t$.
To first order in the timestep, the updated \DF{} is
\begin{equation}
    f_\chi^i \bigg|_{t + \Delta t} = f_\chi^i \bigg|_t + \frac{\dd f_\chi^i}{\dd t} \bigg|_t \, \Delta t ,
    \label{eq:f_chi_dt}
\end{equation}
where we require $f_\chi^i \big|_{t + \Delta t}$ to have the form $n_i (\vec{x}) \, \nu_i' (\vec{v})$, for $\nu_i'$ the new velocity distribution of the particle.

The derivation proceeds from here in detail in \Appx\ref{appx:dm_deriv}.
Briefly, the steps we take are as follows.
We substitute into \Eqn\ref{eq:f_chi_dt} the baryon macroparticle \DF{} $\hat{f}_B$.
This gives a sum over gas particles $j$, each with a different integral over $\dd^3 v_B \, \dd \Omega$.
We simulate this integral by drawing a single Monte Carlo sample independently for every $j$ (denoting the sampled baryon velocity $\vec{v}^{B}_{ij}$ and scattering angles $\Omega_{ij}$).
Then, in order to isolate the velocity distribution on both sides, we integrate both sides over all space.
The result is
\begin{widetext}
\begin{gather}
    \nu_i' (\vec{v})
    =
    \Big( 1 -
    \sum_j P_{ij}
    \Big) \,
    \delta (\vec{v} - \vec{v}_i)
    + \sum_j P_{ij} \,
    \delta \big( \vec{v} - \vec{S}_\chi(\vec{v}_i, \vec{v}_{ij}^{B}, \Omega_{ij}) \big) ,
    \quad \text{where}
    \label{eq:dm-update-vels}
    \\
    P_{ij} \equiv 
    \frac{M_j}{\mu_j} \,
    g_{ij} \,
    \sigma(v_{\rm rel}) \, v_{\rm rel} \, \Delta t ,
    \quad
    g_{ij} \equiv \int \dd^3 x \, W(|\vec{x} - \vec{x}_i|, h_i) \,
    W(|\vec{x} - \vec{x}_j|, h_j) ,
    \quad
    \sigma (v) \equiv \int \dd \Omega \, \frac{\dd\sigma}{\dd\Omega} \bigg|_{v} ,
    \label{eq:dm-update-prob}
\end{gather}
\end{widetext}
and $v_{\rm rel} \equiv |\vec{v}_{i} - \vec{v}_{ij}^{B}|$.
Note the appearance of the total cross section $\sigma (v)$.
We have introduced two intermediate variables to clarify the meaning of this result.
One is $g_{ij}$, introduced by \citet{Rocha+2013-SIDM} for self-interacting dark matter simulations and dubbed the ``geometric factor'' or the ``overlap integral''.
It has dimensions of inverse volume and gives the effective number density of particle $j$ acting on particle $i$.
The other is $P_{ij}$, which is dimensionless and, as long as $P_{ij} \ll 1$, can be understood as a probability.
Indeed, $(M_j / \mu_j) \, g_{ij}$ is the effective number density of microscopic baryons active in this interaction, so $P_{ij}$ has the form $n \, \sigma \, v \, \Delta t$, exactly as one would expect.

The new velocity distribution $\nu_i'$ is a discrete probability distribution:
it gives the post-scattering velocity $\vec{S}_\chi (\vec{v}_i, \vec{v}^{B}_{ij}, \Omega_{ij})$ with probability $P_{ij}$,
or the original velocity $\vec{v}_i$ with probability $1 - \sum_j P_{ij}$.
To pick a new velocity for macroparticle $i$, we can draw a sample from this discrete distribution.
Notice that the time-evolution of the dark matter macroparticles is given by probabilistically scattering them off samples from the baryon velocity distribution in a manner that agrees exactly with the microscopic theory.

\subsection{Baryon evolution}

We next consider the updates to apply to the baryons.
The updated baryon velocity distribution is required to remain Maxwell-Boltzmann.
This suggests an alternative approach in which we directly compute the rates of change of the parameters of the Maxwell-Boltzmann distribution: the bulk velocity and the temperature.
These are
\begin{align}
    n_j (\vec{x}) \, \frac{\dd \vec{V}_j}{\dd t}
    &=
    \int \dd^3 v_B \,
    \frac{\dd f_B^j}{\dd t} \bigg|_{\vec{x}, \vec{v}_B} \,
    \vec{v}_B ,
    \label{eq:defn_dVdt}
    \\
    n_j (\vec{x}) \, \frac{3}{\mu_j} \, \frac{\dd T_j}{\dd t}
    &=
    \int \dd^3 v_B \,
    \frac{\dd f_B^j}{\dd t} \bigg|_{\vec{x}, \vec{v}_B} \,
    (\vec{v}_B - \vec{V}_j)^2 .
    \label{eq:defn_dTdt}
\end{align}
These can be evaluated using techniques very similar to those described by \citet{Ali-Haimoud2019},
augmented by an integral over space (like in the dark matter case).
Details are provided in \Appx\ref{appx:gas_deriv}.
The result is
\begin{widetext}
\begin{align}
    \frac{\dd \vec{V}_j}{\dd t}
    &\approx
    \sum_i
    \frac{M_i}{m_\chi + \mu_j} \, g_{ij} \,
    \mathcal{A} (\vec{v}_i - \vec{V}_j; T_j / \mu_j) \,
    (\vec{v}_i - \vec{V}_j) ,
    \label{eq:gas-dVdt}
    \\
    \frac{3}{2 \, \mu_j} \, \frac{\dd T_j}{\dd t}
    &\approx
    \sum_i \frac{M_i}{m_\chi + \mu_j} \, g_{ij} \, \bigg[
    (\vec{v}_i - \vec{V}_j)^2 \,
    \mathcal{A} (\vec{v}_i - \vec{V}_j; T_j / \mu_j)
    - \frac{\mu_j}{m_\chi + \mu_j} \,
    \mathcal{B} (\vec{v}_i - \vec{V}_j; T_j / \mu_j)
    \bigg] ,
    \label{eq:gas-dTdt}
\end{align}
\end{widetext}
where $g_{ij}$ is the overlap integral as before and $\mathcal{A}$ and $\mathcal{B}$ are defined:
\begin{align}
    \mathcal{A} (\vec{w}; \varsigma^2)
    &\equiv
    \int \dd^3 u \, \mathcal{G} (\vec{w} - \vec{u}, \varsigma^2)
    \, u
    \, \frac{\vec{u} \cdot \vec{w}}{w^2}
    \, \overline{\sigma}(u),
    \label{eq:scrA} \displaybreak[2]
    \\
    \mathcal{B} (\vec{w}; \varsigma^2)
    &\equiv
    \int \dd^3 u \, \mathcal{G} (\vec{w} - \vec{u}, \varsigma^2)
    \, u^3
    \, \overline{\sigma}(u),
    \label{eq:scrB}
    \\
    \overline\sigma (v)
    &=
    \int \dd\Omega \, \frac{\dd\sigma}{\dd\Omega} \bigg|_{v} \,
    (1 - \cos\theta) .
\end{align}
Notice the appearance of the momentum transfer cross section, $\overline{\sigma}$.
If the momentum transfer cross section has a power-law dependence on $v$, then these integrals $\mathcal{A}$ and $\mathcal{B}$ have simple forms in terms of special functions, given in \Appx\ref{appx:gas_deriv}.
In simulation, these rates of change can be treated directly as new sources of acceleration and heating, added onto the total acceleration and heating rates, which are already accumulated for each particle and integrated by the preferred scheme of the simulation.

These rates of change are statistically equal and opposite to the Monte Carlo sampling procedure we have outlined for the dark matter.
Momentum and energy are conserved in expectation value, which we prove in \Appx\ref{appx:equivalence}.

We note that the dark matter and baryon calculations end up using \textit{different} integrals over the differential cross section: 
the dark matter scattering probability depends on the total cross section $\sigma$, while the gas exchange rates depend on the momentum transfer cross section $\overline\sigma$.
In the self-interacting dark matter literature, it is often assumed that one can approximate a non-isotropic cross section in the scattering probability by using isotropic scatterings with the momentum transfer cross section \citep{TulinYu2018}.
For dark matter--baryon interactions, such an assumption must be carefully considered, as it could break the equivalence between the two calculations.

\subsection{GIZMO implementation}

We implement this method in GIZMO \citep{Hopkins2015-GIZMO}, which is based in part on the algorithms of GADGET-2 \citep{Springel2005-GADGET-2}.
We choose to specialize to an isotropic power-law cross section $\sigma (v) = \overline\sigma (v) = \sigma_0 \, (v / c)^n$.
We note that other cross sections, including non-isotropic ones, can be implemented in this framework with relative ease, so long as the integrals defining $\mathcal{A}$ and $\mathcal{B}$ can be computed efficiently.

The core of the implementation is a nearest-neighbors calculation, where dark matter particles seek gas particles and vice versa, during the ``kick'' step of the GIZMO leapfrog integrator.
Each particle searches in a radius equal to twice its softening length.
With this search radius, each pair of particles that overlaps is guaranteed to be found by the neighbor with larger softening length, but may or may not be found by the smaller neighbor.
Accordingly, we must search in both directions (dark matter $\rightarrow$ gas and vice versa); to ensure that each pair is processed only once, we skip a pair whenever the \textit{searching} particle has the smaller softening length.

To process pair $(i, j)$ (taking $i$ to be dark matter, $j$ gas), we first compute the overlap integral $g_{ij}$.
We then compute the contribution of particle $i$ to the acceleration $\dd \vec{V}_j / \dd t$ and heating rate $\dd T_j / \dd t$ of particle $j$, which are accumulated over all pairs that include $j$.
Next, we draw a random sample from the gas velocity distribution $\mathcal{G}(\vec{v} - \vec{V}_j, T_j / \mu_j)$ and use it to compute the scattering probability $P_{ij}$.
If the scattering probability $P_{ij}$ is too large, we reduce the timestep size for the dark matter particle for the next timestep.
The criterion we use is
\begin{equation}
    P_{ij} < \min \bigg(
        0.1,
        10^{-4} \, \bigg[ \frac{m_\chi + \mu_j}{\mu_j} \bigg]^2
    \bigg),
    \label{eq:prob_criterion}
\end{equation}
which requires smaller timesteps if $m_\chi \lesssim \mu_j$ (because in that case scatters are expected to impart a large change on the dark matter velocity).\footnote{%
    Note that the sum $\sum_j P_{ij}$ should also remain small.
    For $m_\chi = 2 \, m_p$, \Eqn\ref{eq:prob_criterion} imposes a maximum of $P_{ij} \lesssim 10^{-3}$, with most probabilities significantly less than this threshold, so this is safely satisfied in our simulations.}
We draw a random number $r$ uniformly between 0 and 1 to decide whether to scatter the dark matter particle.
If we choose to scatter (if $r < P_{ij}$), we choose a new direction $\uvec{v}_{\chi B}'$ isotropically, then compute $\vec{v}_\chi'$ from \Eqn\ref{eq:scatter_chi} and apply it to particle $i$.

We use the default softening kernel, a cubic spline
\begin{equation}
    W(r, h) = \frac{8}{\pi h^3} \, \left\{ \begin{array}{ll}
            1 - 6 \, (\tfrac{r}{h})^2 + 6 \, (\tfrac{r}{h})^3 &
            0 \leq \tfrac{r}{h} \leq \tfrac{1}{2} , \\
            2 \, (1 - \tfrac{r}{h})^3 &
            \tfrac{1}{2} < \tfrac{r}{h} \leq 1 , \\
            0 & \tfrac{r}{h} > 1 .
    \end{array} \right.
\end{equation}
This kernel integrates to 1 and is nonzero for $r \in [0, h)$.
It is designed so that the smoothing is equivalent to smoothing by a Plummer kernel with scale radius $\epsilon = h / 2.8$.

An important detail in the implementation is the computation of the confluent hypergeometric function ${}_1 F_1 (a, b; x)$.
As given in \Appx\ref{appx:gas_deriv}, the integrals $\mathcal{A}$ and $\mathcal{B}$ reduce to simple forms involving ${}_1 F_1$ for our chosen cross section.
We exclusively use this function in the regime of small $a$, small $b$, and negative $x$.
However, the GSL implementation of this function is subject to overflow at modestly large values of $-x$.
We can improve the performance for this specific regime by making use of Kummer's transformation 
and the asymptotic expansion for large values of $x$
\citep[Equations 
\href{https://dlmf.nist.gov/13.2.E39}{13.2.39} and
\href{https://dlmf.nist.gov/13.7.E2}{13.7.2}]{NIST:DLMF}.
Together, these give the following expansion:
\begin{equation}
    {}_1 F_1 (a, b; -x) =
    \frac{\Gamma(b) \, x^{-a}}{\Gamma(b - a)} \,
    \sum_{k=0}^{\infty}
    \frac{(a)_k \, (1 + a - b)_k}{k! \, x^k},
\end{equation}
where $( \cdot )_k$ is the Pochhammer symbol.
In our implementation, we switch to this expansion whenever $|x| \geq 50$ (using the GSL implementation otherwise), truncating the sum at five terms.
This gives fractional errors of less than $10^{-14}$ for all $x$ of interest when compared against the Boost Math library implementation.\footnote{%
    See the Boost library's excellent discussion of the methods they employ to calculate this function accurately:  \url{https://www.boost.org/doc/libs/latest/libs/math/doc/html/math_toolkit/hypergeometric/hypergeometric_1f1.html}.
}

Another important implementation detail is the computation of the overlap integral $g_{ij}$.
It is too expensive to compute as needed at runtime, but it can be tabulated ahead of time for the chosen softening kernel, as it depends only on the distance between the two particles and the ratio of their softening lengths.
This is done via a technique similar to that described by \citet{Fischer+2025}; details are given in \Appx\ref{appx:overlap}.

We note an additional implicit assumption of the method:
that the mean free path length of the particles, $\lambda \sim (n \, \sigma)^{-1}$, must remain larger than the softening lengths of the simulation particles, $h$, in regions where scatterings are important.
If this is violated, then pairs of particles that are farther than $\lambda$ apart can interact, transporting energy over nonphysical distances.
We defer a more comprehensive analysis of this possible error to future work.

Lastly, we note that our implementation properly accounts for the conversion from comoving to physical coordinates and the Hubble flow in GIZMO, although this is not directly tested in the simulations in this paper.

Our implementation is open-source under the GNU General Public License and is publicly available on GitHub.\footnote{\url{https://github.com/cmhainje/gizmo-public}}

\subsection{Comparison to existing methods}
\label{sec:comparison}

The method we use for dark matter is very similar to methods for simulating self-interacting dark matter \citep[e.g.,][]{Rocha+2013-SIDM}.
Our treatment of baryons borrows much technology from analyses of dark matter--baryon interactions in linear cosmology \citep[e.g.,][]{Dvorkin+2014,Ali-Haimoud2019}.
At the time of publication of this article, however, the only existing method specifically designed for simulating dark matter--baryon interactions is that of \citet{Fischer+2025} (\Fischer{}).
We give in this section a detailed comparison of our method to the \Fischer{} scheme for ``rare, large-angle scatters'';
\Fischer{} also gives a scheme for ``frequent, small-angle scatters'' specialized to a certain class of cross section, which largely avoids the pitfalls we describe but is not analogous to our method.

At a high level, \Fischer{} and this work treat dark matter the same way.
In both methods, dark matter particles scatter with neighboring baryon particles with probability $(M_B / m_B) \, g_{ij} \, \sigma \, v \, \Delta t$.
Further, both sample the scattering baryon velocity $\vec{v}_B$ from the Maxwell-Boltzmann distribution described by the bulk velocity and internal energy of the neighbor gas particle.

The key difference is our handling of baryons.
We use the fact that the baryon velocity distribution remains Maxwell-Boltzmann at all times to calculate moments of the Boltzmann equation.
This gives us momentum and heat transfer rates that can be directly applied to the gas macroparticles.
\Fischer{} instead develops a scheme to apply the probabilistic scatters to the baryons.
The issues they must solve, and which we sidestep, are how to conduct a scattering between macroparticles that represent
    (1) different numbers of physical particles (i.e., when $M_\chi / m_\chi \neq M_B / m_B$), and
    (2) different kinds of velocity distributions (i.e., Dirac delta for dark matter and Maxwell-Boltzmann for baryons).
\Fischer{} handles these issues by constructing a ``virtual'' baryon particle for scattering.
The virtual particle has mass $M_{\rm virt}$ set so that $M_{\rm virt} / m_B = M_\chi / m_\chi$, which makes it straightforward to apply a scattering.
After scattering, the changes in momentum and energy of the virtual particle are distributed back to the simulation baryon particle, which makes use of the Maxwell-Boltzmann distribution nature of baryon macroparticles.

The virtual baryon approach has a few key advantages.
One is that it is marginally more efficient: baryons are only updated when a partner dark matter particle is updated, and these updates do not require the calculation of any special functions.
Another is that energy and momentum are conserved on a per-scattering basis.
We contrast these against our code, where
    baryons receive small, smooth updates every timestep,
    these updates require calculating special functions,
    and the split approach means we can only conserve energy and momentum on average.

The primary difficulty with the virtual baryon approach is ensuring that the internal energy of the simulation particle remains positive after the scattering.
This is enforced by rejecting scatterings whenever they result in a negative internal energy.
The rate of these rejections primarily depends on how the mass ratio of the macroparticles, $M_\chi / M_B$, compares to that of the true particles, $m_\chi / m_B$.
If a large fraction of scatters are rejected, the energy transfer between species will be biased.

\begin{figure}
    \centering
    \includegraphics[width=\linewidth]{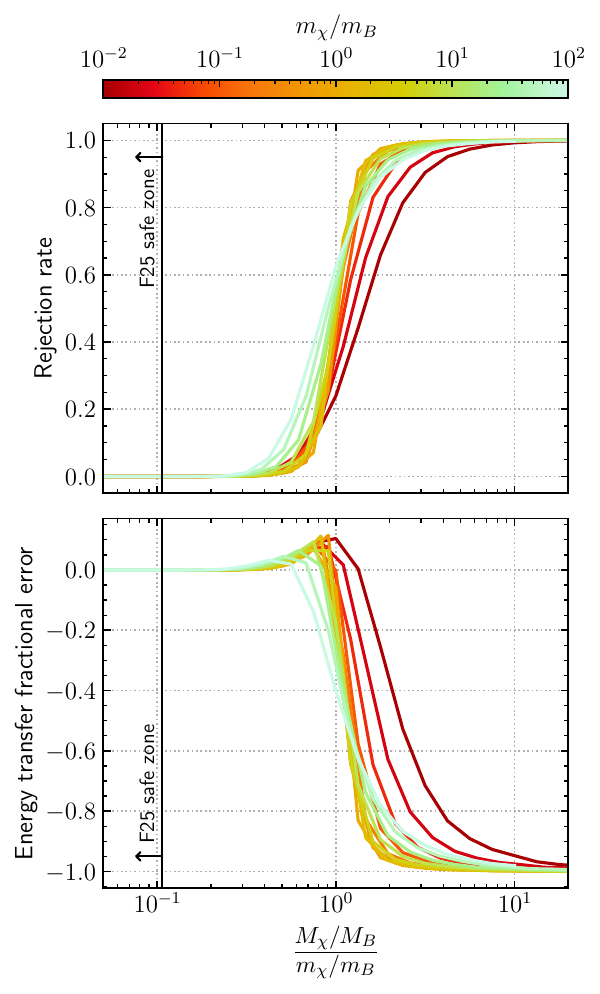}
    \caption{%
    The rejection rate (top) and fractional error of the energy transfer (bottom) for the \citet{Fischer+2025} (\Fischer{}) method as a function of the macroparticle mass ratio $M_\chi / M_B$
    divided by
    the physical particle mass ratio $m_\chi / m_B$.
    The region that is expected to be ``safe'' according to Equation 12 of \Fischer{} is marked with a vertical line and a leftward arrow.
    In agreement with this expectation, we find that the algorithm is valid whenever $M_\chi / M_B \ll m_\chi / m_B$, as the rejection rate is negligible and the energy transfer matches the expectation.
    For $M_\chi / M_B \gtrsim m_\chi / m_B$, the transition to rejecting nearly all scatters happens rather abruptly, leading to a severe underestimate of the energy transfer rate.
    }
    \label{fig:f25-validity}
\end{figure}

We have performed a numerical experiment using the \Fischer{} algorithm to determine what macroparticle mass ratios are valid for a range of true particle mass ratios.
A detailed discussion of our numerical experiment is given in \Appx\ref{appx:f25_validity}.
The rate at which scatters are rejected and the resulting bias of the overall energy transfer are shown in \Fig\ref{fig:f25-validity}.
The vertical bar denotes the upper boundary of the ``safe zone'' given in
Equation 12 of \Fischer{}; one sees that their safe zone is robust and even conservative by a factor of 2, depending on the accuracy required.
In general, the accuracy of the \Fischer{} virtual baryon approach rapidly and severely degrades whenever $M_\chi / M_B \gtrsim \frac{1}{3} \, m_\chi / m_B$.

For small dark matter particle masses, $m_\chi \lesssim m_B$, restricting to the safe zone means that the dark matter macroparticles need much lower mass (i.e., higher resolution) than baryons.
Such a requirement is computationally infeasible in many cosmological and galaxy simulations, where the dark matter is the dominant component of the simulation by mass and high resolution for the baryons is needed.
Because our method makes different trade-offs, it remains accurate even in the $M_\chi / M_B \gtrsim m_\chi / m_B$ regime.

In summary, the \Fischer{} approach has computational advantages, including exact conservation of momentum and energy, which in some regimes make it more efficient than the method we have developed.
However, \Fischer{} is only accurate when the $M_\chi / M_B \lesssim \tfrac{1}{3} \, m_\chi / m_B$.
For small dark matter masses $m_\chi \lesssim m_B$, this restriction would make many cosmological simulations infeasible.
However, our method remains accurate without this restriction, enabling much lower computational cost.
Further, as will be described below, we find nearly identical performance to the \Fischer{} method in a test problem where the \Fischer{} method is accurate.

\section{Tests}
\label{sec:tests}

We have verified our implementation by constructing a test problem that isolates the dark matter--baryon interactions. Explicit details and results are given in \Appx\ref{appx:test_details}; we present a brief summary here.
The test problem consists of a box with periodic boundary conditions filled uniformly with dark matter and gas of specified densities.
The gas particles start at rest and are given some initial internal energy.
The dark matter particles are given velocities sampled from a Maxwell-Boltzmann distribution with some bulk velocity and effective temperature.
Importantly, the dark matter velocity and temperature are not the same as those of the gas, so the system is not initially in equilibrium.
All physics modules in GIZMO are disabled except for dark matter--baryon interactions.
We test a wide variety of initial conditions and mass resolutions for our fiducial dark matter model, which has mass $m_\chi = 2 \, m_p$ and velocity-independent cross section $\sigma = 10 \ {\rm mb}$ ($10^{-26} \ {\rm cm}^2$).
We also test alternative dark matter models for a limited set of initial conditions, including different dark matter masses, $m_\chi \in \{ 10 \, \text{GeV}, 100 \, \text{MeV}, 10 \, \text{MeV} \}$, and velocity-dependent cross sections, $n \in \{-2, -4\}$.

The test problem is designed to facilitate comparison to a direct numerical integration of a system of differential equations derivable under the assumption that the dark matter is Maxwell-Boltzmann distributed at all times.
By this, we do not mean that the dark matter is in equilibrium and unchanging at all times; rather, we mean that the velocity distribution evolves in such a way that it is always of Maxwell-Boltzmann form, with a bulk velocity and temperature that are changing with time.
This assumption does not and should not be expected to hold, as the dark matter departs from a three-dimensional isotropic Gaussian while transferring energy and momentum to the gas.
However, for many choices of the test setup, we find that the dark matter velocity distribution does not stray too far from a thermal distribution, and so the Maxwell-Boltzmann calculation can be used to check the correctness of the implementation.
For other setups where the dark matter evolution departs substantially from a thermal distribution, the Maxwell-Boltzmann prediction becomes invalid and we cannot directly predict the evolution of the system, but we can still verify that the two species come into equilibrium with one another and that momentum and energy are conserved.
A detailed discussion of our test choices, the comparison to the na\"ive Maxwell-Boltzmann prediction, and the results of our simulations are given in \Appx\ref{appx:test_details}.

Coincidentally, our test setup is almost identical to the one employed in \citet{Fischer+2025},
so we re-create the most analogous of their tests (detailed in their Section 3.4).
Our code produces essentially indistinguishable behavior to what they report.

\section{Galaxy simulations}
\label{sec:galaxy}

\begin{table*}
    \centering
    \begin{tabular}[t]{lcc}
        \hline
        \multicolumn{3}{c}{Halo (NFW)} \\
        Virial mass & $M_{200}$ & $1.074 \times 10^{12} \ M_\odot$ \\
        Virial radius & $R_{200}$ & $205.5 \ \text{kpc}$ \\
        Circular velocity & $v_{c,200}$ & $150 \ \text{km}/\text{s}$ \\
        Concentration & $c$ & $10$ \\
        Spin parameter & $\lambda$ & $0.04$ \\
        Particles & $N_h$ & $10^5$ \\ \hline
        \multicolumn{3}{c}{Stellar bulge (Hernquist)} \\
        Mass & $M_b$ & $4.297 \times 10^{9} \ M_\odot$ \\
        Particles & $N_{b}$ & $1.25 \times 10^4$
        \\ \hline
    \end{tabular}
    \begin{tabular}[t]{lcc}
        \hline
        \multicolumn{3}{c}{Disk (exponential)} \\
        Mass & $M_d$ & $4.297 \times 10^{10} \ M_\odot$ \\
        Scale length & $r_d$ & $3.432 \ \text{kpc}$ \\
        Scale height & $z_d$ & $0.3432 \ \text{kpc}$ \\
        Gas fraction & $f_{\text{gas}}$ & 0.2 \\
        Gas temperature & & $10^4 \ \text{K}$ \\
        Gas metal fraction & & $0.02041$ \\
        Stellar particles & $N_{d,*}$ & $10^5$ \\
        Gas particles & $N_{d,\text{gas}}$ & $10^5$ \\ \hline
    \end{tabular}
    \caption{Parameters describing the galaxy initial conditions, from \citet{Kim+2016-AGORA-II}.}
    \label{tab:galaxy_parameters}
\end{table*}

\begin{figure*}
    \centering
    \includegraphics[width=\linewidth]{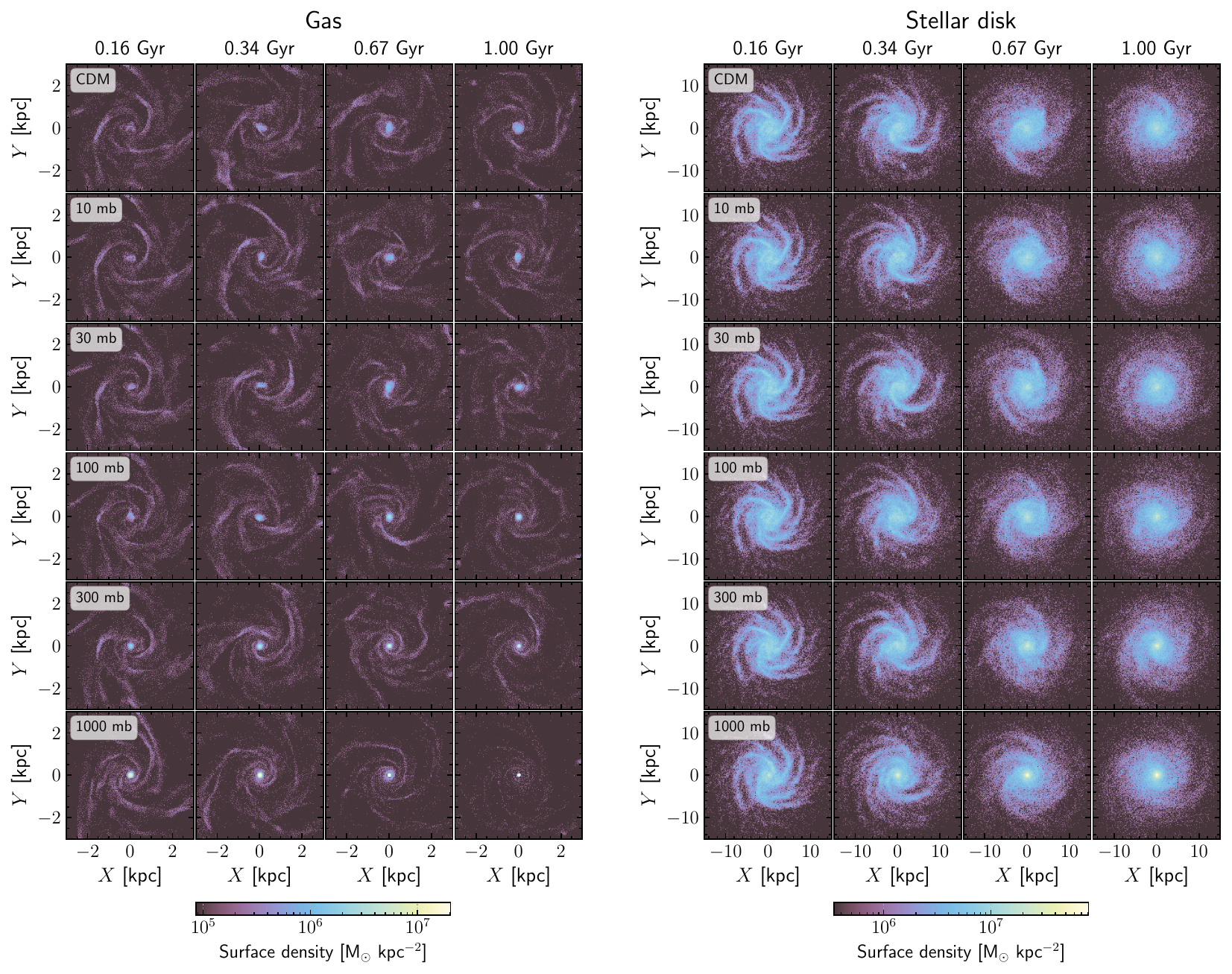}
    \caption{%
    Top-down view of the gas in the center of the disk (left) and the entire stellar disk (right) at a variety of cross sections (increasing from top to bottom) and times (increasing from left to right).
    The two largest cross sections show the development of an extremely compact and dense object in the central gas, as well as the concentration of stars in the center.
    No significant differences are seen between CDM and the smaller cross sections.
    }
    \label{fig:disk}
\end{figure*}

As an initial application of our code to a physical system, we simulate an isolated, $z \sim 1$ Milky Way-like galaxy, composed of an NFW profile dark matter halo of virial mass around $10^{12} \ M_\odot$, an exponential disk of both stars and gas, and a Hernquist profile stellar bulge.
The initial conditions we use are taken from the AGORA isolated disk galaxy test \citep{Kim+2016-AGORA-II}, where the detailed parameters are presented and explained; we give a summary in \Tab\ref{tab:galaxy_parameters}.

With these initial conditions, we perform a suite of simulations using our modified version of GIZMO \citep{Hopkins2015-GIZMO} to include dark matter--baryon interactions.
We choose a dark matter particle mass of $m_\chi = 2 \, m_p$ and velocity-independent cross sections ($n = 0$) of $\{ 0, 10, 30, 100, 300, 1000 \} \ \text{mb}$.
We match the parameters and physics used in \citet{Kim+2016-AGORA-II}, including
    radiative cooling via the Grackle library \citep{Smith+2017-Grackle},
    metal production,
    star formation,
    a simple thermal stellar feedback model,
    a Jeans pressure floor \citep{RobertsonKravtsov2008},
    and Plummer-equivalent gravitational softening lengths of 80 pc for all particle types.
For the kernel used, gravity is exactly Newtonian at distances greater than $2.8$ times the softening length (224 pc in this case).
We denote this length as the ``kernel-support'' softening length, $h_{\rm grav}$, because the softening kernel has support over $r \in [0, h_{\rm grav})$.

Additionally, we use adaptive softening lengths for the gas and dark matter in the calculation of hydrodynamic and DM--b interactions.
These adaptive softening lengths, which we will denote $h_{\rm int}$, can vary down to a floor of 44.8 pc.
(This softening length is the one appearing as $h$ in our derivations.)
The choice of fixed softenings for gravity and adaptive softenings for interactions has precedent:
    \citet{Kim+2016-AGORA-II} makes the same choice for gas, and
    the SIDM implementation in GIZMO \citep{Rocha+2013-SIDM} does the same for dark matter.
In some test simulations, we found that the results of the simulation were not sensitive to changing the dark matter interaction softening length to use a raised floor or to use the same fixed value as $h_{\rm grav}$.

We note that adaptive gravitational softening lengths for dark matter in SIDM simulations have been seen to cause artificial numerical heating \citep[e.g.,][]{Palubski+2024}.
We defer to future work the question of whether adaptive gravitational softenings in our DM--b simulations could have a similar effect.

We evolve the simulations for approximately 1 Gyr.
Since we do not attempt to simulate the cosmic baryon cycle, some of the gas in the galaxy is consumed by star formation, which prematurely slows the dark matter--baryon interactions and therefore sets a limit on the length of the simulation.
While longer simulations would be interesting, especially for smaller cross sections where the impact of interactions takes longer to build up, these would require a model for gas accretion during the evolution of the galaxy.

\subsection{Impacts on morphology}
\label{sec:morph}

The primary effect of the dark matter--baryon interactions is the accumulation of both gas and dark matter in the galaxy center when the cross section is large.
This increased density can be seen by eye in the gas and stellar disk, which are shown face-on at a few times during the simulations in \Fig\ref{fig:disk}.
The gas within the central $\sim 2$ kpc is shown on the left; the entire stellar disk is shown on the right.
Each subfigure shows a grid of images, with the cross section increasing from top to bottom and the time increasing from left to right.

The largest cross sections (300, 1000 mb) show that a considerable amount of gas falls into the center, forming an extremely compact and dense central disk of size $\sim 50 \ {\rm pc}$.
Because the size of this object is substantially smaller than the gravitational softening length, higher resolution will be required to resolve its structure.
For 1000 mb, this process starts sooner and proceeds more quickly.
The stellar disk also concentrates in the innermost $\sim 1 \text{--} 2 \ {\rm kpc}$;
because stars are collisionless, their response demonstrates that the potential changes drastically.
Smaller cross sections show no significant differences from CDM.


\begin{figure}
    \centering
    \includegraphics[width=\linewidth]{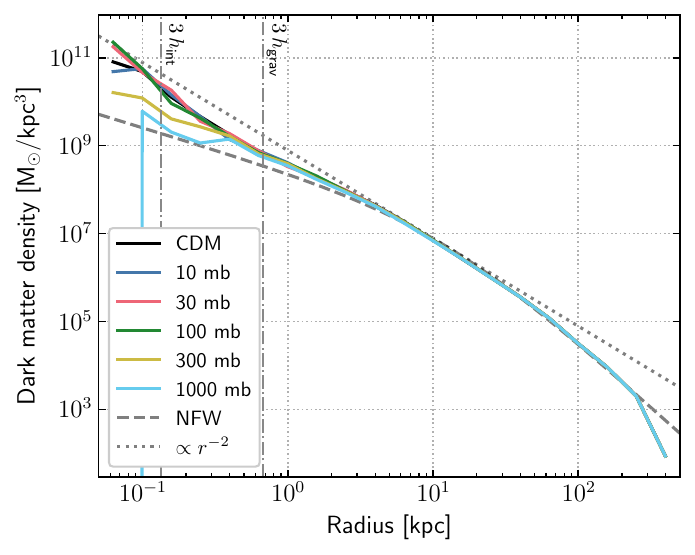}
    \caption{%
    Dark matter density profile at the end of the simulations (at about 1 Gyr).
    Plotted for reference are an NFW profile chosen to match the initial conditions (dashed grey line) and a profile proportional to $r^{-2}$ (dotted grey line).
    All simulations agree for radii beyond a few kpc; differences are only apparent in regions that are gravitationally unresolved in the present simulations.
    }
    \label{fig:dm_dens_profile}
\end{figure}

\begin{figure}
    \centering
    \includegraphics[width=\linewidth]{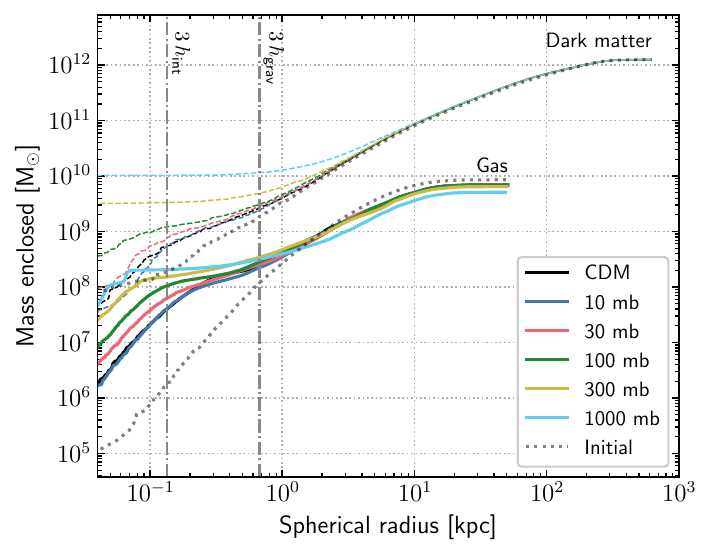}
    \caption{Enclosed dark matter (dashed) and gas (solid) mass profiles at the end of the simulations (at about 1 Gyr).
    Relative to CDM, gas and dark matter concentrate in the center, with the density increasing as a function of cross section.
    }
    \label{fig:encl_mass}
\end{figure}

The accumulation of dark matter into the center can be seen in the final dark matter density profile, plotted in \Fig\ref{fig:dm_dens_profile}.
To facilitate comparison, an NFW profile and a line with slope $r^{-2}$ are shown as well.
We also show vertical lines at three times the minimum interaction and gravitational softening lengths, $h_{\rm int}$ and $h_{\rm grav}$;
features at radii smaller than these lines should be considered unresolved.
All profiles are approximately the same for radii outside of about 1 kpc and closely follow the NFW profile.
The profile is slightly damped for the largest cross sections, which is actually the result of the innermost dark matter being concentrated to radii below the left edge of this figure (cf. \Fig\ref{fig:encl_mass} and discussion below).
However, we caution the reader that this damping is not definitively resolved.


The concentrations of gas and dark matter into the center can also be compared by looking at their enclosed mass profiles, which are shown in \Fig\ref{fig:encl_mass}.
All gas profiles except 1000 mb are roughly the same outside a radius of about 1 kpc; 1000 mb instead shows a flattened profile out to a few kpc and substantially less total gas.
All dark matter profiles appear the same outside of about 5 kpc.
At smaller radii, the gas and dark matter concentrate into the center relative to the case with no DM--b interactions.
At the largest cross sections, this concentration is surrounded by an evacuated region, indicated by horizontal segments of the profile.
This can be seen for the gas in the face-on visualizations of \Fig\ref{fig:disk}.
(This also explains the apparent damping of the density profile in \Fig\ref{fig:dm_dens_profile}.)
The amount of mass that concentrates increases with cross section, and the radius of the concentrated ``object'' decreases with cross section.
While much of the profile of the concentration appears at radii below the gravitational resolution limit, we note that even at $r = 3 \, h_{\rm grav}$ there is a definite trend in the enclosed mass as a function of cross section.
We investigate possible mechanisms driving these inflows---bar formation and transfer of angular momentum between the gas and dark matter---in \Secs\ref{sec:bar} and \ref{sec:angmom}.


\subsection{Impacts on gas and stars}
\label{sec:gas}

\begin{figure}
    \centering
    \includegraphics[width=\linewidth]{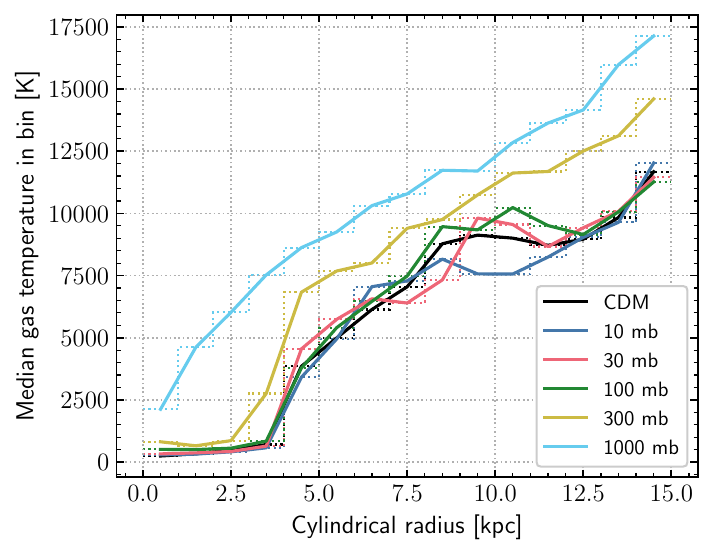}
    \caption{%
    The median gas temperature at the end of the simulations, computed in 1 kpc bins of cylindrical radius.
    Dotted lines indicate the bins in which the temperatures were calculated.
    For cross sections of 300 and 1000 mb, there is a significant temperature increase almost everywhere; no significant differences are seen at smaller cross sections.
    }
    \label{fig:gas_temp_profile}
\end{figure}

Another quantity of interest is the temperature profile of the gas disk.
For the final snapshot of each simulation, we compute the median gas temperature in 1 kpc bins of cylindrical radius up to $R = 15 \ {\rm kpc}$;
the temperature profile is shown in \Fig\ref{fig:gas_temp_profile}.
We see no significant differences at small cross sections.
In the 300 and 1000 mb simulations, however, the gas temperature is increased by a significant amount almost everywhere.
This is expected: in a Milky Way-like galaxy, the dark matter velocity dispersion is of order a few hundred km/s, which translates to an effective temperature of $\mathcal{O}(10^6 \, \text{K})$ for $m_\chi = 2 \, m_p$.
Hence, the dark matter is ``hotter'' than the gas and therefore heats it for sufficiently strong interactions.


\begin{figure}
    \centering
    \includegraphics[width=\linewidth]{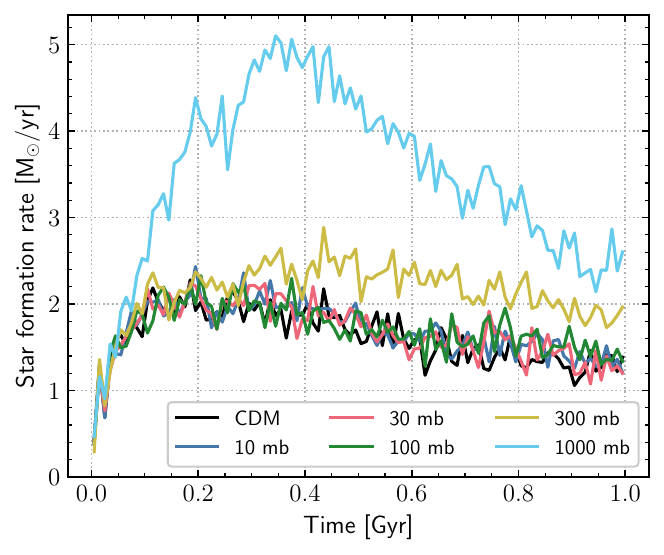}
    \caption{Star formation rate over 1 Gyr for all simulated cross sections.
    For cross sections of 300 and 1000 mb, the star formation rate is increased at nearly all times.
    Smaller cross sections show few significant differences.
    }
    \label{fig:sfr}
\end{figure}

Because dark matter--baryon interactions both heat the gas and change its density, the effects on the star formation history of the galaxy may be non-trivial.
To investigate that, for each snapshot we track the total mass in newly formed stars, and take the change in stellar mass divided by the time interval between snapshots as an estimate of the star formation rate.
\Fig\ref{fig:sfr} shows the evolution of the star formation rate over the history of the simulation.
With cross sections of 300 and 1000 mb, we see enhanced star formation rates starting almost immediately, despite the elevated gas temperature.
At smaller cross sections, there are only small deviations from CDM that are uncorrelated with cross section, and so are likely just stochastic noise.

\subsection{Possible role of bar formation}
\label{sec:bar}

\begin{figure}
    \centering
    \includegraphics[width=\linewidth]{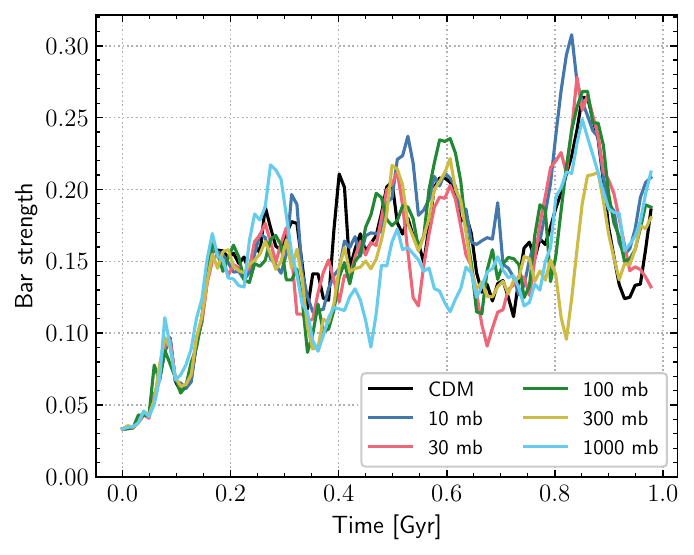}
    \caption{Bar strength of the stellar disk as a function of time.
    The bar strength is computed as the maximum of the second Fourier mode (\Eqn\ref{eq:bar-strength}) over 1 kpc bins of cylindrical radius spanning $R \in [0, 15] \ {\rm kpc}$.
    A strength of 0.1--0.2 is a strong indicator of bar formation.}
    \label{fig:bar_strength}
\end{figure}

Because bar formation is known to drive gas inflows toward the galaxy center \citep{Fanali+2015}, it is important that we disentangle whether dark matter--baryon interactions or a possible bar is responsible for the increased concentration discussed in \Sec\ref{sec:morph}.
To this end, we investigate whether a bar forms in this simulation.
The strength of a bar can be quantified using the second Fourier mode:
\begin{equation}
    b = \frac{
    \big| \sum_j M_j \, \exp (2 i \, \varphi_j) \big|
    }{ \sum_j M_j },
    \label{eq:bar-strength}
\end{equation}
where $M_j$ is the mass of star particle $j$ and $\varphi_j$ is its azimuthal angle \citep{Beane+2023}.
For every snapshot of the simulations, we compute the bar strength $b$ in 1 kpc bins of cylindrical radius up to $R = 15 \ {\rm kpc}$ and record the maximum over the bins.
\Fig\ref{fig:bar_strength} shows the bar strength as a function of time,
revealing that it evolves almost identically across all cross sections, reaching values above 0.15---a strong indicator for bar formation---within 300 Myr.
In particular, during this initial period of bar formation, we find no dependence of either the bar strength or the speed of bar formation on cross section for these initial conditions.

\begin{figure}
    \centering
    \includegraphics[width=\linewidth]{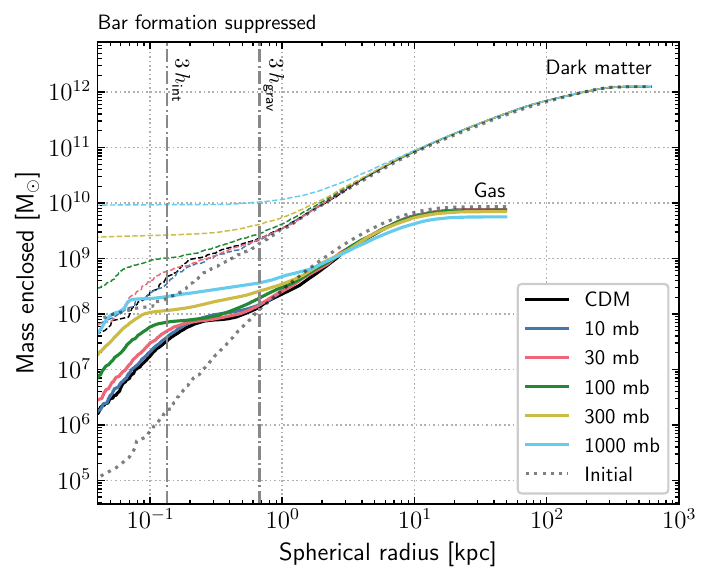}
    \caption{Enclosed dark matter (dashed) and gas (solid) mass profiles at the end of simulations with the stellar disk frozen in order to prevent bar formation.
    These simulations behave similarly to the simulations with bar formation, indicating that the inflow of gas and dark matter to the galaxy center is not the result of bar formation.}
    \label{fig:no_bar}
\end{figure}

Having convinced ourselves that formation of the bar is not affected by the presence or absence of dark matter--baryon interactions, we next isolate the dark matter--baryon effects from bar-driven inflows.
To do this, we perform a set of simulations with the stellar disk frozen in place; since the stars do not move, they cannot form a bar.
The enclosed mass profiles of the end states of these simulations are shown in \Fig\ref{fig:no_bar}.
The results are very similar to the simulations with bar formation, indicating that dark matter--baryon interactions, and not bar formation, are driving gas and dark matter toward the center in this simulation.
We emphasize that this does not mean that dark matter--baryon interactions have no effect on bar formation, and we consider this an interesting question that merits future study.

\subsection{Angular momentum}
\label{sec:angmom}

\begin{figure}
    \centering
    \includegraphics[width=\linewidth]{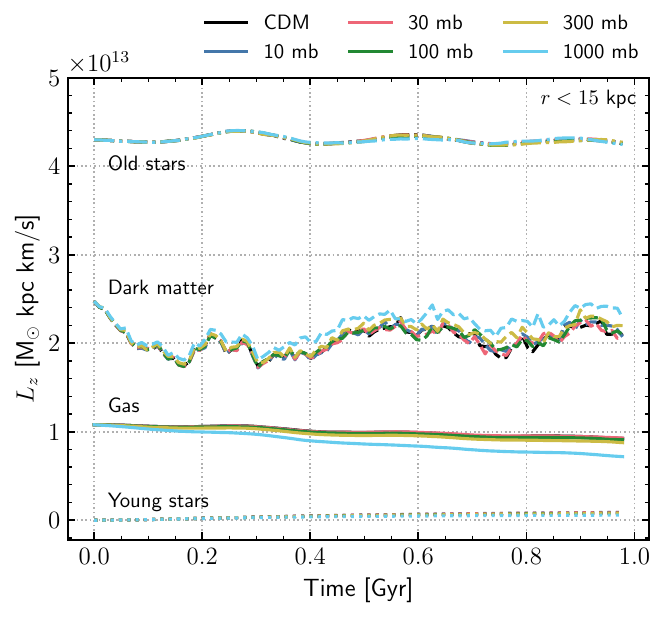}
    \caption{%
    Evolution of the $z$-component of the total angular momentum for particles in the innermost 15 kpc of the galaxy.
    ``Old stars'' denote the stellar disk and bulge that are part of the initial conditions;
    ``young stars'' denote those that are formed out of the gas during the simulation.
    At large cross sections, the dark matter gains angular momentum relative to the CDM simulation, while the gas and young stars lose angular momentum.
    No significant differences are seen between CDM and the smaller cross sections.
    }
    \label{fig:inner_ang_momentum}
\end{figure}

The angular momentum of the gas disk (in particular, its $z$-component) may play a role in the development of gas inflows.
Dark matter--baryon interactions introduce a new possible mechanism for transferring angular momentum into or out of the gas disk.
To explore this, we investigate the evolution of the $z$-component of the angular momentum during these simulations.
Because the dark matter at large radii in general dominates the total angular momentum, we focus on the particles in the innermost 15 kpc.
The evolution of the total angular momentum of each component in this region is shown in \Fig\ref{fig:inner_ang_momentum}.
We see that at large cross section (1000 mb) the dark matter gains angular momentum relative to the CDM simulation, while the gas and newly formed stars lose angular momentum, providing evidence that DM--b interactions are facilitating a transfer of angular momentum.

The direction of the transfer (from gas to dark matter) is sensible.
The gas disk rotates coherently with nonzero average tangential velocity, whereas the dark matter has a nearly isotropic velocity distribution and therefore almost zero average tangential velocity.
Scatters allow the two species to exchange momentum, driving them toward a common bulk velocity, thereby causing the dark matter to pick up tangential velocity and the gas to slow down.


\begin{figure}
    \centering
    \includegraphics[width=\linewidth]{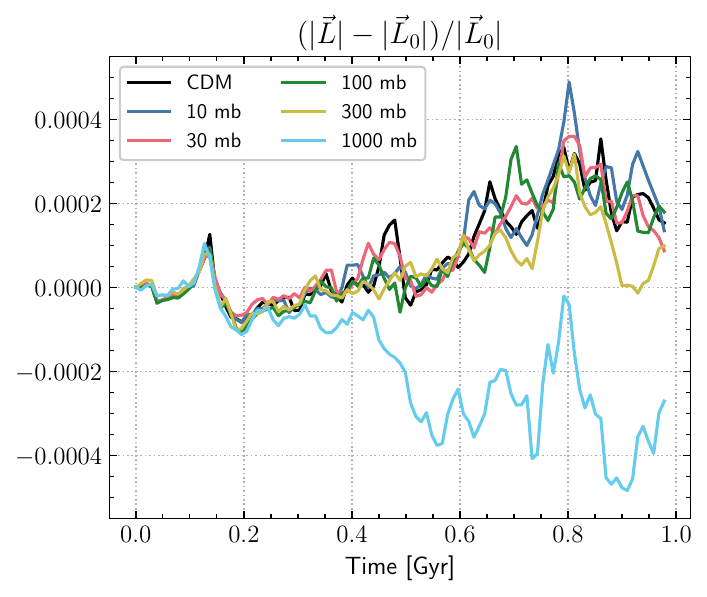}
    \caption{%
    Fractional change in the magnitude of the total angular momentum $\vec{L}$ as a function of time.
    At all cross sections except 1000 mb, we see extremely similar evolution, accumulating about 0.03\% additional angular momentum during the simulation, indicating that there is some conservation error as a baseline.
    At 1000 mb, the accumulation has about the same magnitude but the opposite sign.
    }
    \label{fig:angular-momentum}
\end{figure}

In a galaxy simulation, one cannot directly test for conservation of energy when using dissipative physics like radiative cooling.
However, angular momentum is conserved up to errors from the tree gravity solver, which gives us a useful diagnostic for whether our treatment of the dark matter--baryon interactions has introduced a source of non-conservation.%
\footnote{We note that some small angular momentum non-conservation may be expected due to the finite separation between macroparticles at the time of interactions \citep[as is true for SIDM implementations, see][]{TulinYu2018}, which would naturally be greater with large cross sections due to the increased number of scatterings and exchange rates.}
For each snapshot of our simulations, we compute the total spin angular momentum $\vec{L} = \sum_i M_i \, (\vec{r}_i - \vec{r}_{\rm CM}) \times (\vec{v}_i - \vec{v}_{\rm CM})$ ($\vec{r}_{\rm CM}$ and $\vec{v}_{\rm CM}$ being the position and velocity of the center of mass of the galaxy) and evaluate how the magnitude of this vector evolves.
The evolution is shown in \Fig\ref{fig:angular-momentum}, revealing that there is a nonzero but small change in total angular momentum during the simulation for all cross sections: about 0.03\% per Gyr, even with dark matter--baryon interactions disabled.
Curiously, the \textit{sign} of the non-conservation is reversed for the largest cross section.
This difference is not necessarily concerning;
non-conservation is present even in the CDM simulation, so there is clearly a baseline level of error in solving the CDM physics.
At 1000 mb, the morphology of the galaxy differs greatly from what is seen in the other simulations for much of the simulation, so it is not clear how much of the difference should be attributed directly to non-conservation in scatterings in our implementation versus the evolution of the same baseline error in a strongly reshaped galaxy.
Note in particular that for the first 300 Myr, when the simulations still have similar morphologies, there is no great difference in the total angular momentum at 1000 mb, despite greatly increased exchange rates and scattering rates.

\section{Summary and conclusions}
\label{sec:conclusions}

We have developed a method for incorporating dark matter--baryon interactions in cosmological $N$-body simulations that is accurate even in the challenging domain of dark matter with mass comparable to or lower than that of the baryon.
We have implemented the method for isotropic power-law cross sections of the form $\sigma = \sigma_0 \, (v / c)^n$ in the simulation code GIZMO \citep{Hopkins2015-GIZMO}.
Our implementation is open-source, released under the GNU General Public License, and available on GitHub.\footnote{\url{https://github.com/cmhainje/gizmo-public}}
We tested the accuracy of the model in a range of environments, including one test that is reproduced from \citet{Fischer+2025} where it gives virtually identical behavior.

We have applied our implementation to the evolution of an isolated Milky Way-like disk galaxy, using initial conditions from the AGORA project \citep{Kim+2016-AGORA-II} and simulating the system for 1 Gyr.
We considered a dark matter particle of mass $2 \, m_p$ with isotropic, velocity-independent DM--b cross sections of 0, 10, 30, 100, 300, and 1000 mb.
For the cases of 300 and 1000 mb, the DM--b interactions clearly change the mass distribution in the central kiloparsec of the galaxy, significantly increasing the concentration of baryons and dark matter.
This occurs both with and without bar formation.
Interactions at this strength also heat the gas, increase the star formation rate, and facilitate the transfer of angular momentum out of the gas disk.

In our simulations with cross sections of 100 mb and less, the DM--b interactions do not have enough time to have any significant impact.
This is not surprising, since the initial gas supply is partially used up, and we do not simulate gas accretion into the galaxy.
Properly simulating the effect of a DM--b cross section of order 10 mb or less demands high-quality hydrodynamics, a model of gas accretion, and a detailed study of the fidelity of the subgrid models of stellar feedback.
These efforts are not justified in this exploratory investigation of an idealized, isolated galaxy.
In future work, we plan to study cosmological galaxy formation in the interesting cross section range, with the required sophisticated treatment of the ``gastrophysics''.

The code and method we have developed and presented here to incorporate dark matter--baryon interactions provide a practical, efficient, and reliable tool for the community.
It should enable robust limits to be placed on various scenarios for dark matter--baryon interactions based on properties of galaxies.

\begin{acknowledgments}
We thank
    Michael Blanton,
    Shy Genel,
    Chris Hayward,
    David W.~Hogg,
    Clark Miyamoto,
and Ben Wandelt
for helpful discussions,
    and the anonymous referee for useful comments and suggestions.
We thank Moritz Fischer for his helpful feedback on our discussion of the method he and his co-authors developed.
We especially thank Angus Beane for his advice and debugging aid,
and Phil Hopkins for illuminating conversations about GIZMO and the possibilities for implementing this new feature.
CH is supported by the National Science Foundation Graduate Research Fellowship under Grant Number DGE-2234660.
This research of GRF was supported in part by the Simons Foundation.
This work made use of New York University IT High Performance Computing resources, services, and staff expertise.
\end{acknowledgments}

\bibliography{main}{}
\bibliographystyle{aasjournalv7}

\appendix

\section{Derivation of dark matter evolution}
\label{appx:dm_deriv}

We begin with the Boltzmann equation for dark matter particle $i$.
As written in \Eqn\ref{eq:boltzmann-individual}, the single-particle \DF{} satisfies
\begin{equation}
    \frac{\dd f_{\chi}^i}{\dd t} (\vec{v}_\chi)
    =
    \int \dd^3 v_B \, \dd \Omega \, 
    \frac{\dd \sigma}{\dd \Omega} \,
    v_{\chi B} \,
    \big[
        f_{\chi}^i (\vec{S}_\chi) \, \hat{f}_B (\vec{S}_B)
        - f_{\chi}^i (\vec{v}_\chi) \, \hat{f}_B (\vec{v}_B)
    \big] .
\end{equation}
The Boltzmann equation is made of the difference of two rates: the rate of scatters \textit{into} a state with velocity $\vec{v}_\chi$ and the rate of scatters \textit{out of} velocity $\vec{v}_\chi$.
The rate of scatters ``in'' can be conveniently reformulated by instead integrating over all possible initial velocities $\vec{v}_\chi', \vec{v}_B'$ and scattering angles $\Omega$, conditioned on the requirement that the post-scatter velocity is $\vec{v}_\chi$:
\begin{equation}
    \frac{\dd f_{\chi}^i}{\dd t} (\vec{v}_\chi)
    =
    \int \dd^3 v_\chi' \, \dd^3 v_B' \, \dd \Omega \, 
    \frac{\dd \sigma}{\dd \Omega} \,
    v_{\chi' B'} \,
    f_{\chi}^i (\vec{v}_\chi') \, \hat{f}_B (\vec{v}_B') \,
    \delta \big( \vec{S}_\chi(\vec{v}_\chi') - \vec{v}_\chi \big)
    -
    \int \dd^3 v_B \, \dd \Omega \, 
    \frac{\dd \sigma}{\dd \Omega} \,
    v_{\chi B} \,
    f_{\chi}^i (\vec{v}_\chi) \, \hat{f}_B (\vec{v}_B) .
\end{equation}
We will primarily use this form of the Boltzmann equation in the following derivations.

Consider the evolution of this single-particle \DF{} over a timestep $\Delta t$.
To first order in the timestep, the update is
\begin{equation}
    f_\chi^i \bigg|_{t + \Delta t} = f_\chi^i \bigg|_t + \frac{\dd f_\chi^i}{\dd t} \bigg|_t \, \Delta t .
\end{equation}
The updated \DF{} has the form $n_i(\vec{x}) \, \nu_i' (\vec{v})$, where $\nu_i'$ is the new velocity distribution of the particle.
Plugging in $\hat{f}_B$, we thus find
\begin{align}
    n_i (\vec{x}) \, \nu_i' (\vec{v})
    &=
    n_i (\vec{x}) \,
    \bigg[
    \bigg( 1 -
    \sum_j n_j (\vec{x})
    \int \dd^3 v_B \,
    \dd \Omega \,
    \mathcal{G}(\vec{v}_B - \vec{V}_j, T_j / \mu_j) \,
    \frac{\dd\sigma}{\dd\Omega} \,
    v_{iB} \, \Delta t
    \bigg) \,
    \delta (\vec{v} - \vec{v}_i)
    \\ \nonumber
    &\qquad\qquad\qquad
    + \sum_j n_j (\vec{x})
    \int \dd^3 v_B \,
    \dd \Omega \,
    \mathcal{G}(\vec{v}_B - \vec{V}_j, T_j / \mu_j) \,
    \frac{\dd\sigma}{\dd\Omega} \,
    v_{iB} \, \Delta t \,
    \delta(\vec{v} - \vec{S}_\chi(\vec{v}_i))
    \bigg] .
\end{align}
The result contains two integrals over $\dd^3 v_B$ for each gas particle $j$.
The second is tricky, as the integration variables $\vec{v}_B$ and $\Omega$ are used in the scattering transformation $\vec{S}_\chi$ inside the Dirac delta function.
Therefore, it is expedient to approximate the integrals instead by Monte Carlo sampling.
The integrals can be performed independently for each particle $j$ using single samples from the baryon Maxwell-Boltzmann distribution, which we denote $\vec{v}^{B}_{ij}$, and the angular component of the differential cross section, $\Omega_{ij}$.
The result is
\begin{align}
    n_i (\vec{x}) \, \nu_i' (\vec{v})
    =
    n_i (\vec{x}) \,
    \bigg[
    &\bigg( 1 -
    \sum_j
    n_j (\vec{x}) \,
    \sigma(v_{\rm rel}) \, v_{\rm rel} \, \Delta t
    \bigg) \,
    \delta (\vec{v} - \vec{v}_i)
    \\ \nonumber
    &
    + \sum_j
    n_j (\vec{x}) \,
    \sigma(v_{\rm rel}) \, v_{\rm rel} \, \Delta t \,
    \delta(\vec{v} - \vec{S}_\chi(\vec{v}_i; \vec{v}^{B}_{ij}, \Omega_{ij}))
    \bigg] ,
\end{align}
where $v_{\rm rel} \equiv |\vec{v}_i - \vec{v}^{B}_{ij}|$, and $\sigma(v)$ is the total cross section, given by $\sigma (v) = \int \dd \Omega \, (\dd\sigma / \dd\Omega) \big|_{v}$.
The total cross section appears because one can decompose the differential cross section as
\begin{equation}
    \frac{\dd\sigma}{\dd\Omega} \bigg|_{v, \Omega}
    =
    \sigma(v) \, p(\Omega \,|\, v),
\end{equation}
where $p(\Omega \,|\, v)$ is the probability density of scattering angles conditioned on the relative velocity.
The sample $\Omega_{ij}$ is drawn from $p(\Omega \,|\, v_{\rm rel})$, allowing us to approximate $\dd\sigma/\dd\Omega \approx \sigma(v) \, \delta(\Omega - \Omega_{ij})$, and the rest follows.

In order to isolate $\nu_i' (\vec{v})$ from all $\vec{x}$ dependence, we can integrate both sides over all space.
We find
\begin{equation}
    \nu_i' (\vec{v}) =
    \Big( 1 - \sum_j P_{ij} \Big) \,
    \delta (\vec{v} - \vec{v}_i)
    + \sum_j P_{ij} \,
    \delta \big( \vec{v} - \vec{S}_\chi(\vec{v}_i, \vec{v}^{B}_{ij}, \Omega_{ij}) \big) ,
\end{equation}
where we have introduced
\begin{equation}
    g_{ij} \equiv \int \dd^3 x \, W(|\vec{x} - \vec{x}_i|, h_i) \,
    W(|\vec{x} - \vec{x}_j|, h_j)
    \quad \text{and} \quad
    P_{ij} \equiv 
    \frac{M_j}{\mu_j} \,
    g_{ij} \,
    \sigma(v_{\rm rel}) \, v_{\rm rel} \, \Delta t
    .
\end{equation}

\section{Derivation of baryon evolution}
\label{appx:gas_deriv}

Because the baryons are assumed to thermalize on timescales faster than the timesteps of the simulation, we impose that the baryon velocity distribution is at all times Maxwell-Boltzmann.
Thus, the evolution of gas particle $j$ is described by the rates of change of the bulk velocity and temperature:
\begin{align}
    n_j (\vec{x}) \,
    \frac{\dd \vec{V}_j}{\dd t}
    &= \int \dd^3 v_B \,
    \frac{\dd f_B^j}{\dd t} \bigg|_{\vec{x}, \vec{v}_B} \,
    \vec{v}_B ,
    \\
    n_j (\vec{x}) \,
    \frac{3}{\mu_j} \,
    \frac{\dd T_j}{\dd t}
    &= \int \dd^3 v_B \,
    \frac{\dd f_B^j}{\dd t} \bigg|_{\vec{x}, \vec{v}_B} \,
    (\vec{v}_B - \vec{V}_j)^2 .
\end{align}

As we proceed, we will use the notation
$ f_\chi = f_\chi(\vec{x}, \vec{v}_\chi) $,
$ f_\chi' = f_\chi(\vec{x}, \vec{v}_\chi') $,
$ f_B = f_B^j (\vec{x}, \vec{v}_B) $,
$ f_B' = f_B^j (\vec{x}, \vec{v}_B') $.
We start with the acceleration.
Inserting the Boltzmann equation gives
\begin{align}
    n_j (\vec{x}) \,
    \frac{\dd \vec{V}_j}{\dd t}
    &=
    \int \dd^3 v_B \,
    \dd^3 v_\chi' \, \dd^3 v_B' \, \dd \Omega \,
    f_\chi' \, f_B' \, \frac{\dd\sigma}{\dd\Omega} \,
    v_{B'\chi'} \,
    \delta \big( \vec{v}_B - \vec{S}(\vec{v}_B') \big) \,
    \vec{v}_B
    \nonumber \\ &\qquad
    - \int \dd^3 v_B \, \dd^3 v_\chi \, \dd\Omega \,
        f_\chi \, f_B \, \frac{\dd\sigma}{\dd\Omega} \, v_{B\chi} \, \vec{v}_B .
\end{align}
The Dirac delta function in the first integral can be integrated immediately.
Then, the two terms are identical up to a relabeling of primed and unprimed velocities and the replacement of $\vec{v}_B$ by $\vec{S}(\vec{v}_B)$.
Using these facts, we find
\begin{align}
    n_j (\vec{x}) \, \frac{\dd \vec{V}_j}{\dd t}
    &=
    \int \dd^3 v_\chi \, \dd^3 v_B \, \dd \Omega \,
    f_\chi \, f_B \, \frac{\dd\sigma}{\dd\Omega} \,
    v_{B\chi} \, \big[ \vec{S}(\vec{v}_B) - \vec{v}_B \big]
    \\ &=
    \int \dd^3 v_\chi \, \dd^3 v_B \, \dd \Omega \,
    f_\chi \, f_B \, \frac{\dd\sigma}{\dd\Omega} \,
    v_{B\chi}^2 \, \frac{m_\chi}{m_\chi + \mu_j} \,
    \big[ \uvec{v}_{B\chi}' - \uvec{v}_{B\chi} \big] .
    \label{eq:baryon_dVdt_integral}
\end{align}
For fixed $\vec{v}_\chi, \vec{v}_B$, the integral over $\Omega$ can only result in a vector aligned with $\uvec{v}_{B\chi}$, so we can project these vectors along this direction.
In other words, we can make the following substitution in the integrand:
$
\uvec{v}_{B\chi}' - \uvec{v}_{B\chi}
\mapsto
\uvec{v}_{B\chi} \, \big[ \uvec{v}_{B\chi} \cdot \uvec{v}_{B\chi}' - 1 \big]
$.
We now introduce the momentum transfer cross section:
\begin{equation}
    \overline\sigma (v_{\rm rel})
    \equiv
    \int \dd\Omega \, \frac{\dd\sigma}{\dd\Omega} \bigg|_{v_{\rm rel}}
    \, [ 1 - \cos\theta ]
    =
    \int \dd\Omega \, \frac{\dd\sigma}{\dd\Omega} \bigg|_{v_{\rm rel}}
    \, [ 1 - \uvec{v}_{\rm rel}' \cdot \uvec{v}_{\rm rel} ].
\end{equation}
Thus,
\begin{align}
    n_j (\vec{x}) \, \frac{\dd \vec{V}_j}{\dd t} &=
    - \frac{m_\chi}{m_\chi + \mu_j}
    \int \dd^3 v_\chi \, \dd^3 v_B \, 
    f_\chi \, f_B \, \overline{\sigma}(v_{B\chi}) \, v_{B\chi} \, \vec{v}_{B\chi} .
\end{align}
Now we insert the Maxwell-Boltzmann form of $f_B^j$.
Introducing $\vec{u} \equiv \vec{v}_\chi - \vec{v}_B$ and $\vec{w} \equiv \vec{v}_\chi - \vec{V}_j$, we arrive at
\begin{align}
    n_j (\vec{x}) \, \frac{\dd \vec{V}_j}{\dd t}
    &=
    \frac{m_\chi}{m_\chi + \mu_j} \, n_j (\vec{x})
    \int \dd^3 v_\chi \, f_\chi \,
    \bigg[ \underbrace{
        \int \dd^3 u \, \mathcal{G}(\vec{w} - \vec{u}, T_j / \mu_j) \,
        \overline{\sigma}(u) \, u \, \frac{\vec{u} \cdot \vec{w}}{w^2}
    }_{\equiv \mathcal{A}(\vec{w}; T_j / \mu_j)} \bigg]
    \, \vec{w} ,
\end{align}
Finally we can insert the \DF{} for the dark matter and integrate both sides over space:
\begin{align}
    n_j (\vec{x}) \, \frac{\dd \vec{V}_j}{\dd t}
    &\approx
    \frac{m_\chi}{m_\chi + \mu_j} \, n_j (\vec{x})
    \sum_i n_i (\vec{x}) \,
    \mathcal{A}(\vec{v}_i - \vec{V}_j; T_j / \mu_j) \,
    (\vec{v}_i - \vec{V}_j)
    \\
    \implies \frac{\dd \vec{V}_j}{\dd t}
    &=
    \sum_i
    \frac{M_i}{m_\chi + \mu_j} \, g_{ij} \,
    \mathcal{A} (\vec{v}_i - \vec{V}_j; T_j / \mu_j) \,
    (\vec{v}_i - \vec{V}_j) .
\end{align}

The heating rate proceeds similarly.
We begin by inserting the Boltzmann equation:
\begin{align}
    n_j (\vec{x}) \,
    \frac{3}{\mu_j} \,
    \frac{\dd T_j}{\dd t}
    &=
    \begin{aligned}[t]
    &\int \dd^3 v_B \,
    \dd^3 v_\chi' \, \dd^3 v_B' \, \dd \Omega \,
    f_\chi' \, f_B' \, \frac{\dd\sigma}{\dd\Omega} \,
    v_{B'\chi'} \,
    \delta \big( \vec{v}_B - \vec{S}(\vec{v}_B') \big) \,
    (\vec{v}_B - \vec{V}_j)^2
    \\
    &\qquad
    - \int \dd^3 v_B \, \dd^3 v_\chi \, \dd\Omega \,
        f_\chi \, f_B \, \frac{\dd\sigma}{\dd\Omega} \, v_{B\chi} \, (\vec{v}_B - \vec{V}_j)^2
    \end{aligned}
    \\ &=
    \int \dd^3 v_\chi \, \dd^3 v_B \, \dd \Omega \,
    f_\chi \, f_B \, \frac{\dd\sigma}{\dd\Omega} \,
    v_{B\chi} \,
    \big[
    ( \vec{S}(\vec{v}_B) - \vec{V}_j )^2
    - (\vec{v}_B - \vec{V}_j)^2
    \big]
    \\ &=
    \int \dd^3 v_\chi \, \dd^3 v_B \, \dd \Omega \,
    f_\chi \, f_B \, \frac{\dd\sigma}{\dd\Omega} \,
    v_{B\chi} \,
    \big[
    ( \vec{S}(\vec{v}_B) - \vec{v}_B )^2
    + 2 \, (\vec{v}_B - \vec{V}_j) \cdot (\vec{S}(\vec{v}_B) - \vec{v}_B)
    \big] .
\end{align}
We now split this into two terms.
The first contains
\begin{align}
    \int \dd \Omega \,
    \frac{\dd\sigma}{\dd\Omega} \,
    ( \vec{S}(\vec{v}_B) - \vec{v}_B )^2
    =
    \bigg( \frac{m_\chi \, v_{B\chi}}{m_\chi + \mu_j} \bigg)^2
    \int \dd \Omega \,
    \frac{\dd\sigma}{\dd\Omega} \,
    ( \uvec{v}_{B\chi}' - \uvec{v}_{B\chi} )^2
    =
    2 \, \bigg( \frac{m_\chi \, v_{B\chi}}{m_\chi + \mu_j} \bigg)^2
    \overline{\sigma} (v_{B\chi}) .
\end{align}
The second:
\begin{align}
    (\vec{v}_B - \vec{V}_j) \cdot 
    \int \dd \Omega \,
    \frac{\dd\sigma}{\dd\Omega} \,
    (\vec{S}(\vec{v}_B) - \vec{v}_B)
    &=
    -\frac{m_\chi}{m_\chi + \mu_j} \,
    \overline{\sigma} (v_{B\chi}) \, v_{B\chi} \, 
    (\vec{v}_B - \vec{V}_j) \cdot  \vec{v}_{B\chi}
    \\ &=
    -\frac{m_\chi}{m_\chi + \mu_j} \,
    \overline{\sigma} (v_{B\chi}) \, 
    v_{B \chi} \,
    \big[
    v_{B\chi}^2 \, 
    +
    (\vec{v}_\chi - \vec{V}_j) \cdot  \vec{v}_{B\chi}
    \big] .
\end{align}
Putting these in,
\begin{align}
    n_j (\vec{x}) \,
    \frac{3}{\mu_j} \,
    \frac{\dd T_j}{\dd t}
    &=
    2 \, \frac{m_\chi}{m_\chi + \mu_j} \,
    \int \dd^3 v_\chi \, \dd^3 v_B \, 
    f_\chi \, f_B \, 
    \overline\sigma (v_{B\chi}) \,
    \bigg[
    \frac{m_\chi}{m_\chi + \mu_j} \, v_{B\chi}^3
    - v_{B\chi}^3
    - v_{B\chi} \, (\vec{v}_\chi - \vec{V}_j) \cdot \vec{v}_{B\chi}
    \bigg]
    \\ &=
    2 \, \frac{m_\chi}{m_\chi + \mu_j} \,
    \int \dd^3 v_\chi \, \dd^3 v_B \, 
    f_\chi \, f_B \, 
    \overline\sigma (v_{B\chi}) \,
    \bigg[
    - \frac{\mu_j}{m_\chi + \mu_j} \, v_{B\chi}^3
    - v_{B\chi} \, (\vec{v}_\chi - \vec{V}_j) \cdot \vec{v}_{B\chi}
    \bigg] .
\end{align}
Now, like before, we insert the Maxwell-Boltzmann form of $f_B^j$, change variables, and find
\begin{align}
    n_j (\vec{x}) \,
    \frac{3}{\mu_j} \,
    \frac{\dd T_j}{\dd t}
    &=
    2 \, \frac{m_\chi}{m_\chi + \mu_j} \, n_j (\vec{x})
    \int \dd^3 v_\chi \,
    f_\chi \,
    \bigg[
    - \frac{\mu_j}{m_\chi + \mu_j} \, \mathcal{B}(\vec{w}; T_j/\mu_j)
    + (\vec{v}_\chi - \vec{V}_j)^2 \, \mathcal{A}(\vec{w}; T_j/\mu_j)
    \bigg] ,
\end{align}
where we introduced
\begin{equation}
    \mathcal{B} (\vec{w}; \varsigma^2)
    \equiv
    \int \dd^3 u \, \mathcal{G} (\vec{w} - \vec{u}, \varsigma^2)
    \, u^3
    \, \overline{\sigma}(u).
\end{equation}
Finally we can insert the form of $f_\chi$ and integrate both sides over space:
\begin{align}
    n_j (\vec{x}) \,
    \frac{3}{2 \, \mu_j} \,
    \frac{\dd T_j}{\dd t}
    &=
    \frac{m_\chi}{m_\chi + \mu_j} \, n_j (\vec{x}) \,
    \sum_i n_i (\vec{x}) \, 
    \bigg[
    (\vec{v}_i - \vec{V}_j)^2 \, \mathcal{A}(\vec{v}_i - \vec{V}_j; T_j/\mu_j)
    - \frac{\mu_j}{m_\chi + \mu_j} \, \mathcal{B}(\vec{v}_i - \vec{V}_j; T_j/\mu_j)
    \bigg]
    \\
    \implies
    \frac{3}{2 \, \mu_j} \,
    \frac{\dd T_j}{\dd t}
    &=
    \sum_i \frac{M_i}{m_\chi + \mu_j} \, g_{ij}
    \bigg[
    (\vec{v}_i - \vec{V}_j)^2 \, \mathcal{A}(\vec{v}_i - \vec{V}_j; T_j/\mu_j)
    - \frac{\mu_j}{m_\chi + \mu_j} \, \mathcal{B}(\vec{v}_i - \vec{V}_j; T_j/\mu_j)
    \bigg] .
\end{align}

As noted by \citet{Ali-Haimoud2019}, $\mathcal{A}$ and $\mathcal{B}$ have simple forms in terms of special functions if the momentum transfer cross section has a power-law dependence on $v$.
Specifically, if $\overline\sigma(v_{\rm rel}) = \sigma_0 \, (v_{\rm rel} / c)^n$ for integer $n \geq -4$, then
\begin{align}
    \mathcal{A}(w; \varsigma^2)
    &=
    \frac{\alpha_n}{3} \, \frac{\sigma_0}{c^n}
    \, \varsigma^{n + 1}
    \, {}_1 F_1 \left(
        -\tfrac{n+1}{2},
        \tfrac{5}{2},
        -\tfrac{w^2}{2 \, \varsigma^2} 
    \right),
    \\
    \mathcal{B}(w; \varsigma^2)
    &=
    \alpha_n \, \frac{\sigma_0}{c^n}
    \, \varsigma^{n+3}
    \, {}_1 F_1 \left(
        -\tfrac{n+3}{2},
        \tfrac{3}{2},
        -\tfrac{w^2}{2 \, \varsigma^2}
    \right),
    \\
    \alpha_n &\equiv \frac{2^{(n + 5)/2} \, \Gamma(3 + \tfrac{n}{2})}{\sqrt{\pi}},
\end{align}
where ${}_1 F_1$ is the confluent hypergeometric function of the first kind.

\section{Equivalence of the two approaches}
\label{appx:equivalence}

In this appendix, we show that momentum and energy are conserved in expectation value given the two approaches, so the two approaches are statistically equivalent.
For dark matter, we have in the end computed the updated velocity distribution of particle $i$ to be
\begin{equation}
    \nu_i' (\vec{v}) \bigg|_{t + \Delta t}
    =
    \frac{m_\chi}{M_i} \,
    \int \dd^3 x \, \bigg(
    f_\chi^i \bigg|_{t}
    +
    \frac{\dd f_\chi^i}{\dd t} \bigg|_{t} \, \Delta t
    \bigg) .
\end{equation}
At the end of a timestep, we draw a sample from this distribution $\nu_i'$, which itself involves approximating $\dd f_\chi^i / \dd t$ by drawing samples from the distributions of baryon velocities and scattering angles.
The average change in velocity, then, is given by integrating over all possible samples (i.e., reverting the sampling approximations):
\begin{align}
    \mathbb{E}[\Delta \vec{v}_i]
    &=
    \frac{m_\chi}{M_i} \,
    \int \dd^3 x \, \dd^3 v \, \Bigg(
    f_\chi^i \bigg|_{t}
    +
    \frac{\dd f_\chi^i}{\dd t} \bigg|_{t} \, \Delta t
    \Bigg) \, \vec{v} - \vec{v}_i
    \nonumber \\ &=
    \frac{m_\chi}{M_i} \, \int \dd^3 x \, \dd^3 v \,
    \frac{\dd f_\chi^i}{\dd t} \bigg|_{t} \, \Delta t \,
    \vec{v} .
\end{align}
Notice that the integral over $(\dd f/\dd t) \, \vec{v}$ is exactly how we define the mean acceleration for the baryons.
One can follow the derivation of $\dd \vec{V}_j / \dd t$ in \Appx\ref{appx:gas_deriv} line-by-line with the subscripts $B$ and $\chi$ exchanged to obtain this expectation value;
the entire derivation is symmetric.
The result is
\begin{align}
    \mathbb{E}[\Delta \vec{v}_i]
    &=
    \sum_j
    \frac{M_j}{m_\chi + \mu_j} \, g_{ij} \,
    \mathcal{A} (\vec{v}_i - \vec{V}_j; T_j / \mu_j) \,
    (\vec{V}_j - \vec{v}_i) 
    \, \Delta t.
\end{align}

The change in energy can be similarly considered.
\begin{align}
    \mathbb{E}[\Delta E_i]
    &=
    \frac{M_i}{2} \,
    \frac{m_\chi}{M_i} \,
    \int \dd^3 x \, \dd^3 v \, \Bigg(
    f_\chi^i \bigg|_{t}
    +
    \frac{\dd f_\chi^i}{\dd t} \bigg|_{t} \, \Delta t
    \Bigg) \, \vec{v}^2 - \vec{v}_i^2
    \\ &=
    \frac{m_\chi}{2} \,
    \int \dd^3 x \, \dd^3 v \,
    \frac{\dd f_\chi^i}{\dd t} \bigg|_{t} \, \Delta t \,
    \vec{v}^2
    \\ &=
    \frac{m_\chi}{2} \,
    \int \dd^3 x \, \dd^3 v \,
    \dd^3 v_B \, \dd \Omega \, f_\chi \, f_B \,
    \frac{\dd\sigma}{\dd\Omega} \, v_{\chi B} \,
    \big[ \vec{S}(\vec{v})^2 - \vec{v}^2 \big]
    \, \Delta t
    \\ &=
    \frac{m_\chi}{2} \,
    \int \dd^3 x \, \dd^3 v \,
    \dd^3 v_B \, \dd \Omega \, f_\chi \, f_B \,
    \frac{\dd\sigma}{\dd\Omega} \, v_{\chi B} \,
    \big[ (\vec{S}(\vec{v}) - \vec{v})^2 + 2 \, \vec{v} \cdot (\vec{S}(\vec{v}) - \vec{v}) \big]
    \, \Delta t
    \\ &=
    \frac{m_\chi \, \mu_j}{m_\chi + \mu_j}
    \int \dd^3 x \, 
    \dd^3 v \,
    \dd^3 v_B \, f_\chi \, f_B \,
    \overline\sigma \, \bigg[
    - \frac{m_\chi}{m_\chi + \mu_j} \, v_{\chi B}^3
    - v_{\chi B} \, \vec{v}_B \cdot \vec{v}_{\chi B}
    \bigg]
    \, \Delta t
    \\ &=
    \frac{m_\chi \, \mu_j}{m_\chi + \mu_j}
    \int \dd^3 x \, 
    \dd^3 v \,
    \dd^3 v_B \, f_\chi \, f_B \,
    \overline\sigma \, \bigg[
    \frac{\mu_j}{m_\chi + \mu_j} \, v_{\chi B}^3
    - v_{\chi B} \, \vec{v}_\chi \cdot \vec{v}_{\chi B}
    \bigg]
    \, \Delta t
    \\ &=
    m_\chi \,
    \sum_j
    \frac{M_j}{m_\chi + \mu_j} \,
    n_j (\vec{x}) \,
    \int \dd^3 x \, 
    \dd^3 v \,
    f_\chi \, \bigg[
    \frac{\mu_j}{m_\chi + \mu_j} \, \mathcal{B} (\vec{w}; T_j / \mu_j)
    - \vec{v}_\chi \cdot (\vec{v}_\chi - \vec{V}_j) \, \mathcal{A} (\vec{w}; T_j / \mu_j)
    \bigg]
    \, \Delta t
    \\ &=
    M_i \,
    \sum_j
    \frac{M_j}{m_\chi + \mu_j} \,
    g_{ij} \,
    \bigg[
    \frac{\mu_j}{m_\chi + \mu_j} \, \mathcal{B} (\vec{v}_i - \vec{V}_j; T_j / \mu_j)
    - \vec{v}_i \cdot (\vec{v}_i - \vec{V}_j) \, \mathcal{A} (\vec{v}_i - \vec{V}_j; T_j / \mu_j)
    \bigg]
    \, \Delta t .
\end{align}

Suppose we have exactly one dark matter particle $i$ and one gas particle $j$.
Then, the rates of change satisfy the following relationships:
\begin{gather}
    M_i \, \mathbb{E}[\Delta \vec{v}_i]
    + M_j \, \frac{\dd \vec{V}_j}{\dd t} \, \Delta t
    = 0 ,
    \\
    \mathbb{E}[\Delta E_i]
    +
    \frac{3 \, M_j}{2 \, \mu_j} \,
    \frac{\dd T_j}{\dd t} \, \Delta t
    +
    M_j \, \vec{V}_j \cdot \frac{\dd \vec{V}_j}{\dd t} \, \Delta t
    = 0 .
\end{gather}
These relationships are recognizable as conservation of the expectation values of energy and momentum.

As we extend to many dark matter and many gas particles, these relationships continue to hold for each individual pair.
This is because both the full expectation values we have just computed \textit{and} the baryonic rates of change simply sum over neighbors of the other type.
Hence, for each pair $(i, j)$, equal and opposite terms appear in the expectation values for the change to $i$ and in the rates of change of $j$.

\section{Computation of overlap integral}
\label{appx:overlap}

Let us consider the computation of the overlap factor $g_{ij}$.
It is defined
\begin{equation}
    g_{ij} \equiv
    \int \dd^3 x \,
    W(|\vec{x} - \vec{x}_i|; h_i) \,
    W(|\vec{x} - \vec{x}_j|; h_j) .
\end{equation}
We assume that the gas and dark matter are softened with the same kernel $W$;
the transformations and tabulation described below will need to be reworked if the two use different softening kernels.

Define the distance $|\vec{x}_i - \vec{x}_j| \equiv d$.
Let us translate our integration variable to place $\vec{x}_i$ at the origin.
The integral is invariant to rotation, so we choose to place $\vec{x}_j$ at $d \, \uvec{z}$.
Then,
\begin{equation}
    g_{ij} \equiv
    \int \dd^3 x \,
    W(|\vec{x}|; h_i) \,
    W(|\vec{x} - d \, \uvec{z}|; h_j) .
\end{equation}

We now move to spherical coordinates: $\vec{x} \to (r, \theta, \phi)$,
so
\begin{align}
    g_{ij} 
    &=
    \int \dd r \, \dd c \, \dd \phi \, r^2
    W(r; h_i) \,
    W(\sqrt{r^2 + d^2 - 2 \, d \, r \, c}; h_j)
    \\ &=
    2 \pi \,
    \int_{0}^{h_i} \dd r \, r^2 \,
    W(r; h_i) \,
    \int_{-1}^{+1} \dd c \, 
    W(\sqrt{r^2 + d^2 - 2 \, d \, r \, c}; h_j) ,
\end{align}
with $c \equiv \cos\theta$.

Generally, the softening kernel depends only on the ratio of its arguments, up to a normalizing factor in terms of $h$.
In other words,
\begin{equation}
    W(x; h) = \frac{1}{h^3} \, w(x / h) ,
\end{equation}
for some $w$.
Let $s \equiv r / h_i$ and $\delta \equiv d / h_i$. Then,
\begin{align}
    g_{ij} 
    &=
    2 \pi \int \dd r \, \dd c \, r^2 \,
    W(r; h_i) \,
    W(\sqrt{r^2 + d^2 - 2 \, d \, r \, c}; h_j)
    \\ &=
    \frac{2 \pi}{h_i^3 \, h_j^3}
    \int \dd r \, \dd c \, r^2 \,
    w(r / h_i) \,
    w(\sqrt{r^2 + d^2 - 2 \, d \, r \, c} / h_j)
    \\ &=
    \frac{2 \pi}{h_j^3} \,
    \int_{0}^{1} \dd s \, s^2 \,
    w(s) \,
    \int_{-1}^{+1} \dd c \, 
    w(\sqrt{s^2 + \delta^2 - 2 \, \delta \, s \, c} / (h_j / h_i)) .
\end{align}
For a given softening kernel, the integral that remains can be tabulated in terms of two parameters: $\delta$ and $h_j / h_i$.

Because $g_{ij}$ is symmetric, we need only tabulate values of the ratio $h_j / h_i$ between 0 and 1;
any ratio $h_j / h_i > 1$ can be handled by swapping $i \leftrightarrow j$.

The range of $\delta$ can also be considered.
We assume $W$ is zero beyond the smoothing length, so the overlap is zero if $d > h_i + h_j$.
Thus, the range of $\delta$ that gives non-zero overlap can be found:
\begin{equation}
    d < h_i + h_j
    \implies
    \delta < 1 + \frac{h_j}{h_i} .
\end{equation}
Since we choose $i$ and $j$ such that $h_j / h_i \leq 1$,
$\delta$ only needs to be tabulated from $0$ to $2$.
For small values of $h_j / h_i$, parts of the table will simply be zero.

We note that this method is almost the same as that described by \citet{Fischer+2025}, although we use spherical rather than cylindrical coordinates to better match the SIDM implementation in GIZMO \citep{Rocha+2013-SIDM}.

\section{Regime of validity of the \Fischer{} method}
\label{appx:f25_validity}

As discussed in \Sec\ref{sec:comparison}, we have performed a numerical experiment to quantify the regions of validity of the ``rare, large-angle scattering'' method of \citet[][\Fischer{}]{Fischer+2025}.
Here we describe that experiment.

We consider two macroparticles, one dark matter and one baryon, with velocities like those of our test method (see \Sec\ref{sec:tests} or the more detailed discussion below).
For this single pair of macroparticles, assumed to be overlapping with some fiducial value for $g_{ij}$, we perform 100\,000 scatterings.
Each scattering involves randomly sampling
    a dark matter velocity from a Maxwell-Boltzmann distribution,
    a ``virtual baryon'' velocity from a Maxwell-Boltzmann distribution, and
    a scattering angle.
We then compute the change to the bulk velocity $\vec{V}_B$ and the internal specific energy $U_B$ of the baryon according to the algorithm described in \Fischer{}.

We are primarily interested in the energy transfer.
We define the total specific energy of the baryon to be
\begin{equation}
    \mathcal{E}_B = \frac{1}{2} \vec{V}_B^2 + U_B .
\end{equation}
In each trial $i$, the scattering occurs with some probability $P_i$, which depends on the sampled velocities.
So, the \textit{expected} change in the total energy of the baryon for trial $i$ is
\begin{equation}
    \Delta\mathcal{E}_i
    =
    P_i \, (\mathcal{E}_B' - \mathcal{E}_B)
    +
    (1 - P_i) \, 0
    =
    P_i \, (\mathcal{E}_B' - \mathcal{E}_B) .
\end{equation}
By averaging this quantity over many trials, we compute the expectation value of the change in energy, integrating by Monte Carlo over scattering angle and the dark matter and baryon velocity distributions:
\begin{equation}
    \mathbb{E}[ \Delta\mathcal{E}_B ]
    =
    \frac{1}{N_{\text{trials}}} \,
    \sum_{i \in \text{trials}}
    P_i \, (\mathcal{E}_B' - \mathcal{E}_B) .
\end{equation}

In the \Fischer{} algorithm, a scattering is rejected if it results in negative internal energy $U_B'$.
Hence, given all these trial scatterings, the expected change in energy that will actually result from the \Fischer{} algorithm is given by averaging over only those trials where $U_B' > 0$:
\begin{equation}
    \mathbb{E}[ \Delta\mathcal{E}_B ]_{\text{actual}}
    =
    \frac{1}{N_{\text{trials} | U_B' > 0}} \,
    \sum_{i \in \text{trials} | U_B' > 0}
    P_i \, (\mathcal{E}_B' - \mathcal{E}_B) .
\end{equation}
We can also compute the expected rejection rate, which is the probability-weighted fraction of scatterings that were rejected:
\begin{equation}
    f_{\text{reject}}
    =
    \frac
        {\sum_{i \in \text{trials}} P_i \, \mathbf{1}(U_B' \leq 0)}
        {\sum_{i \in \text{trials}} P_i} ,
\end{equation}
where $\mathbf{1}(\text{cond})$ is the condition function: 1 if the condition is true and 0 otherwise.

\begin{figure}
    \centering
    \includegraphics[width=0.45\linewidth]{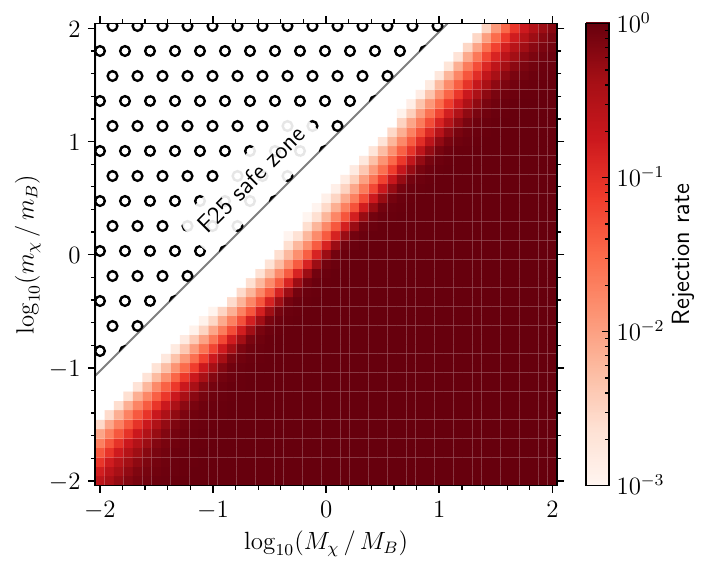}
    \includegraphics[width=0.45\linewidth]{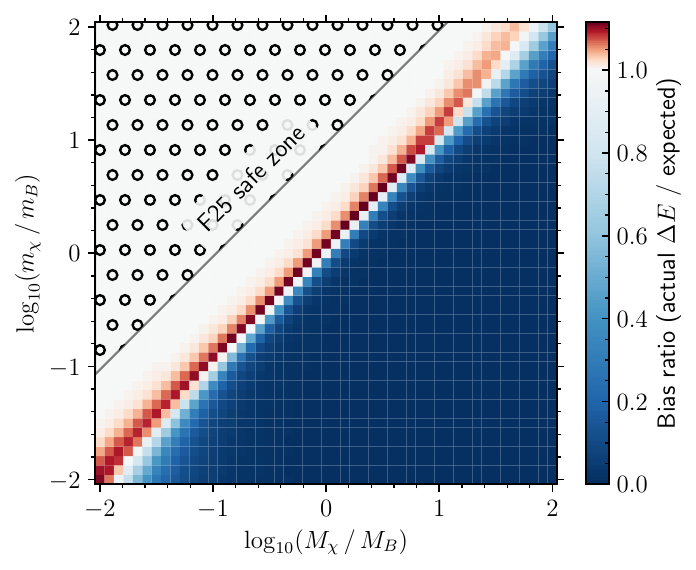}
    \caption{
    The expected rejection rate (left) and energy transfer bias ratio (right) of the \citet{Fischer+2025} algorithm as a function of the mass ratios of the true particles, $m_\chi / m_B$, and of the macroparticles, $M_\chi / M_B$.
    The accuracy of the method is highly degraded whenever $M_\chi / M_B \gtrsim m_\chi / m_B$.
    An analytically predicted ``safe zone'' \citep[Eq.~12,][]{Fischer+2025} is shown hatched with circles.
    }
    \label{fig:f25-imshows}
\end{figure}

We expect the rejection rate to depend primarily on two mass ratios:
    that of the macroparticles $M_\chi / M_B$
    and of the physical particles $m_\chi / m_B$.
We therefore conduct this experiment for a wide range of both ratios.
The resulting rejection rates and energy change bias fractions (actual divided by expected) are shown as heatmaps in \Fig\ref{fig:f25-imshows}.
Also shown on these figures is the expected ``safe zone'' described in \Fischer{}:
\begin{equation}
    \frac{M_\chi}{M_B} < \frac{3}{28} \, \frac{m_\chi}{m_B} .
\end{equation}

We find that the rejection rate is zero and the bias fraction is one everywhere in the expected safe zone, and even for $M_\chi / M_B$ slightly greater than the safe zone threshold.
However, performance degrades steeply once $M_\chi / M_B \gtrsim m_\chi / m_B$, with the rejection rate becoming 100\% almost everywhere.
In general, the effect of the rejections is to reduce the number of scatters and therefore to bias the energy transfer downward, which is borne out by the observed bias ratio.
In fact, for most of the mass ratios tested, the bias ratio becomes exactly zero, with energy transfer completely shut off.
However, there is a small region where the rejections actually enhance the energy transfer;
this is because the distribution of energy transfers is broad when the macroparticle mass ratio is similar to that of the physical particles, and the rejection step only cuts off the negative tail of the distribution.

\section{Test details}
\label{appx:test_details}

As described in \Sec\ref{sec:tests}, we test our implementation with a simple but fairly generic test problem.
We set up a $(10 \ {\rm kpc})^3$ box with periodic boundary conditions and fill it uniformly with gas and dark matter of densities $\rho_B, \rho_\chi$.
The gas cells are initialized at rest on a regular grid with internal energy set to $(3/2) \, (T_B^0 / m_B)$, for $T_B^0$ an initial temperature and $m_B$ the baryon particle mass, assumed to be constant and equal to $m_p$.
The dark matter particles are given uniformly random positions and velocities sampled from a Maxwell-Boltzmann distribution with mean $\vec{V}_\chi = V_\chi^0 \, \uvec{x}$ and temperature $T_\chi^0$.
We use $N_B$ gas cells and $N_\chi$ dark matter particles.
We then simulate the system with all physics disabled except for dark matter--baryon interactions.

We perform this test for many choices of these quantities, detailed in \Tab\ref{tab:test_params}.
The standard setup we choose is
    $\rho_B = \rho_\chi = 10^{10} \ M_\odot/{\rm kpc}^{3}$,
    $T_B^0 = 10^4 \ {\rm K}$,
    $T_\chi^0 = 10^6 \ {\rm K}$, and
    $V_\chi^0 = 200 \ {\rm km}/{\rm s}$,
which is similar to the conditions near the center of our isolated disk galaxy.
The simulations are then analyzed by computing the total momentum (along $x$), the temperature, and the energy of each component for each snapshot.

\begin{table}[h]
    \centering
    \setlength{\tabcolsep}{1em}
    \begin{tabular}{lcccccccccc}
    \hline
        Name
        & $m_\chi$
        & $\sigma_0$
        & $n$
        & $V_\chi^0$
        & $T_\chi^0$
        & $T_B^0$
        & $\rho_\chi$
        & $\rho_B$
        & $N_\chi$
        & $N_B$
    \\
        {}
        & {}
        & {\footnotesize ${\rm cm}^{2}$}
        & {}
        & {\footnotesize ${\rm km}/{\rm s}$}
        & \multicolumn{2}{c}{\footnotesize K} 
        & \multicolumn{2}{c}{\footnotesize $M_\odot / {\rm kpc}^{3}$} 
        & {}
        & {}
    \\ \hline
Fiducial & $2 \, m_p$ & $10^{-26}$ & $0$ & $200$ & $10^{6}$ & $10^{4}$ & $10^{10}$ & $10^{10}$ & $32^3$ & $32^3$ \\
$V_\chi = 0$ & $2 \, m_p$ & $10^{-26}$ & $0$ & $\bm{0}$ & $10^{6}$ & $10^{4}$ & $10^{10}$ & $10^{10}$ & $32^3$ & $32^3$ \\
$T_\chi = T_B = 0$ & $2 \, m_p$ & $10^{-26}$ & $0$ & $200$ & $\bm{10^{1}}$ & $\bm{10^{1}}$ & $10^{10}$ & $10^{10}$ & $32^3$ & $32^3$ \\
$\rho_B = 10 \, \rho_\chi$ & $2 \, m_p$ & $10^{-26}$ & $0$ & $200$ & $10^{6}$ & $10^{4}$ & $\bm{10^{9}}$ & $10^{10}$ & $32^3$ & $32^3$ \\
$\rho_\chi = 10 \, \rho_B$ & $2 \, m_p$ & $10^{-26}$ & $0$ & $200$ & $10^{6}$ & $10^{4}$ & $10^{10}$ & $\bm{10^{9}}$ & $32^3$ & $32^3$ \\
$N_\chi = 24^3$ & $2 \, m_p$ & $10^{-26}$ & $0$ & $200$ & $10^{6}$ & $10^{4}$ & $10^{10}$ & $10^{10}$ & $\bm{24^3}$ & $32^3$ \\
$N_\chi = N_B = 24^3$ & $2 \, m_p$ & $10^{-26}$ & $0$ & $200$ & $10^{6}$ & $10^{4}$ & $10^{10}$ & $10^{10}$ & $\bm{24^3}$ & $\bm{24^3}$ \\
$m_\chi = 10 \ {\rm GeV}$ & $\bm{10 \ {\rm GeV}}$ & $10^{-26}$ & $0$ & $200$ & $\bm{5 \times 10^{6}}$ & $10^{4}$ & $10^{10}$ & $10^{10}$ & $32^3$ & $32^3$ \\
$m_\chi = 100 \ {\rm MeV}$ & $\bm{100 \ {\rm MeV}}$ & $\bm{2 \times 10^{-27}}$ & $0$ & $200$ & $\bm{5 \times 10^{4}}$ & $10^{4}$ & $10^{10}$ & $10^{10}$ & $32^3$ & $32^3$ \\
$m_\chi = 10 \ {\rm MeV}$ & $\bm{10 \ {\rm MeV}}$ & $\bm{2 \times 10^{-27}}$ & $0$ & $200$ & $\bm{5 \times 10^{3}}$ & $10^{4}$ & $10^{10}$ & $10^{10}$ & $32^3$ & $32^3$ \\
$\sigma \sim v^{-2}$ & $2 \, m_p$ & $\bm{2 \times 10^{-33}}$ & $\bm{-2}$ & $200$ & $10^{6}$ & $10^{4}$ & $10^{10}$ & $10^{10}$ & $32^3$ & $32^3$ \\
$\sigma \sim v^{-4}$ & $2 \, m_p$ & $\bm{5 \times 10^{-40}}$ & $\bm{-4}$ & $200$ & $10^{6}$ & $10^{4}$ & $10^{10}$ & $10^{10}$ & $32^3$ & $32^3$ \\
F25 \S 3.4
    & $\bm{m_p}$
    & $\bm{1.6726 \times 10^{-23}}$
    & $0$
    & $\bm{0}$
    & $\bm{484.6}$
    & $\bm{72.69}$
    & $\bm{10^7}$
    & $\bm{10^7}$
    & $\bm{10^5}$
    & $\bm{21^3}$
\\
    \hline
    \end{tabular}
    \caption{%
    Parameter choices for tests;
    those that differ from the ``fiducial'' test are indicated in bold.
    }
    \label{tab:test_params}
\end{table}

A simple numerical prediction for this problem can be made if the dark matter has a Maxwell-Boltzmann distribution at all times.
Under this assumption, each population is described only by its bulk velocity and temperature, and the rates of change of these quantities are determined by a simple system of differential equations, resulting from taking the first two moments of the Boltzmann equation.
(See \citet{Ali-Haimoud2019} for details.)
In this particular test problem, the fact that both species have zero bulk velocity along $y$ and $z$ allows us to simplify the system to the following:
\begin{align}
    \frac{\dd V_\chi}{\dd t} &= \frac{\rho_B}{m_\chi + m_B} \, \mathcal{A}(V_{\chi B}; v_{\rm th}^2) \, (V_B - V_\chi),
    \\
    \frac{\dd V_B}{\dd t} &= \frac{\rho_\chi}{m_\chi + m_B} \, \mathcal{A}(V_{\chi B}; v_{\rm th}^2) \, (V_\chi - V_B),
    \\
    \frac{\dd T_\chi}{\dd t} &= \frac{\rho_B}{(m_\chi + m_B) \, v_{\rm th}^2} \left[ 
    \mathcal{B}(V_{\chi B}; v_{\rm th}^2) \, \frac{T_B - T_\chi}{m_\chi + m_B}
    + \mathcal{A}(V_{\chi B}; v_{\rm th}^2) \, \frac{T_\chi}{m_\chi} \, V_{\chi B}^2
    \right],
    \\
    \frac{\dd T_B}{\dd t} &= \frac{\rho_\chi}{(m_\chi + m_B) \, v_{\rm th}^2} \left[ 
    \mathcal{B}(V_{\chi B}; v_{\rm th}^2) \, \frac{T_\chi - T_B}{m_\chi + m_B}
    + \mathcal{A}(V_{\chi B}; v_{\rm th}^2) \, \frac{T_B}{m_B} \, V_{\chi B}^2
    \right],
\end{align}
where
$V_{\chi B} = |V_\chi - V_B|$,
$v_{\rm th}^2 = T_B / m_B + T_\chi / m_\chi$,
and $\mathcal{A}$ and $\mathcal{B}$ are given in \Eqns\ref{eq:scrA} and~\ref{eq:scrB}.
Given choices for 
    the densities $\rho_B, \rho_\chi$,
    the dark matter mass $m_\chi$,
    the cross section parameters $\sigma_0, n$,
    and the initial velocity distribution parameters $V_\chi^0, T_B^0, T_\chi^0$,
this system of equations can be straightforwardly numerically integrated.
We call the result of this integration the ``na\"ive Maxwell-Boltzmann prediction.''

The assumption that the dark matter distribution is at all times Maxwell-Boltzmann does not (and should not) hold for the system in this test problem.
However, as long as the distribution of dark matter velocities does not depart \textit{too} far from Gaussian, we see very similar behavior between the simulation and the na\"ive Maxwell-Boltzmann prediction.
To quantify the departure from Maxwell-Boltzmann of the dark matter velocity distribution, we use the Jarque-Bera statistic \citep{Jarque-Bera1980}:
\begin{equation}
    {\rm JB}
    =
    \frac{1}{6} \, \bigg[
        S^2 + \frac{1}{4} \,
        \big( K - 3 \big)^2
    \bigg],
\end{equation}
where $S$ is the skewness and $K$ is the kurtosis of the velocities.
(Note that typically this is multiplied by the number of particles, which we forgo in order to compare simulations at different resolutions.)
Large values of ${\rm JB}$ mean large departures from Maxwell-Boltzmann.
For the analysis of our simulations, we compute ${\rm JB}$ for the dark matter velocities independently along $x$, $y$, and $z$ for each snapshot.
We then take the maximum value seen at any time along any axis as our guide for whether the Maxwell-Boltzmann assumption is satisfactory for a given test.

We should make explicit what ``failure'' for these tests would look like.
For a simulation where the JB statistic is small at all times, the temperatures and bulk velocities must remain close to the Maxwell-Boltzmann prediction at all times.
Departure from the prediction would indicate a failure of the implementation.
For a simulation where the JB statistic becomes large, the Maxwell-Boltzmann prediction can no longer be trusted.
Then, we have no ground truth to compare against, but we could still see failure if momentum and energy are not conserved, if the two species do not come into equilibrium, and if the evolution of the momentum and temperature is not smooth and sensible.

\begin{figure}
    \centering
    \includegraphics[width=\linewidth]{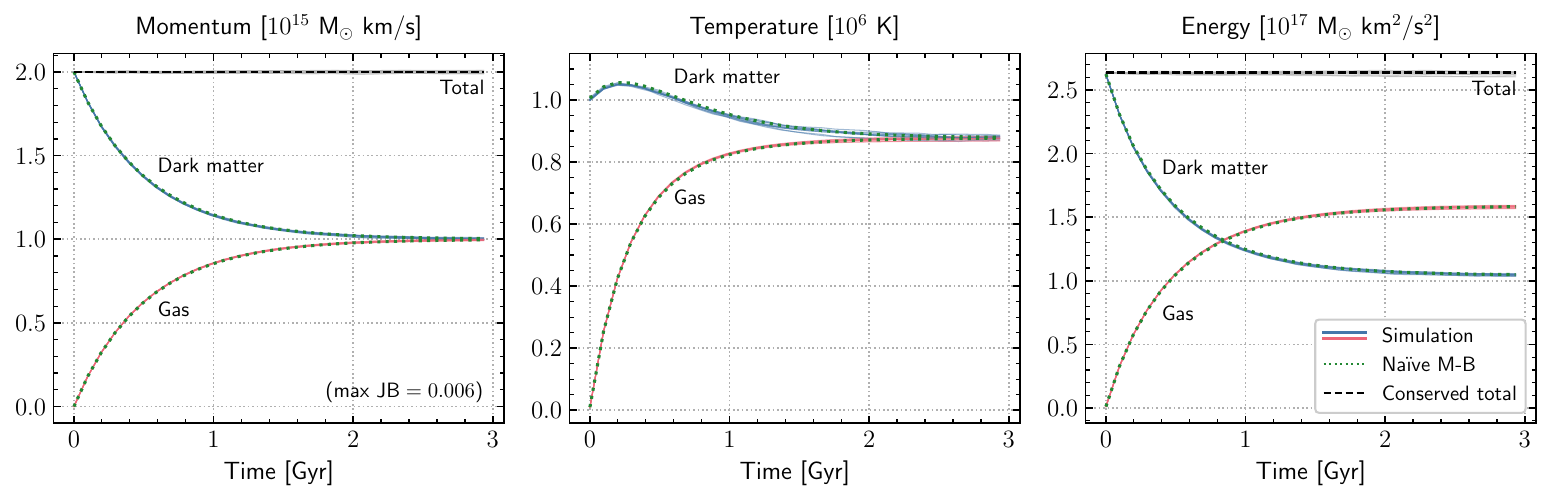}
    \caption{%
    Global momentum along $x$ (left), temperature (middle), and energy (right) of the components of our test simulations (solid colored) as a function of time.
    A black dashed line shows the initial value of the total momentum and energy in order to easily see non-conservation.
    A green dotted line shows a na\"ive analytic prediction for the evolution of the system based on an assumption that the dark matter distribution is always Maxwell-Boltzmann.
    The maximum value of the Jarque-Bera statistic is given (``max JB'') as an indicator of how non-thermal the dark matter distribution becomes during the test; larger values indicate larger deviations from Maxwell-Boltzmann.
    This figure shows the ``fiducial'' test, run with ten different random seeds.
    All tests individually show good agreement both with each other and with the M-B prediction, although some scatter is visible due to the stochastic treatment of scatterings in the dark matter.
    We note that the scatter is larger in the temperature and energy than the momentum, and does not appear to greatly affect the resulting gas temperatures.
    }
    \label{fig:tests-fiducial}
\end{figure}

\begin{figure}
    \centering
    \includegraphics[width=0.83\linewidth]{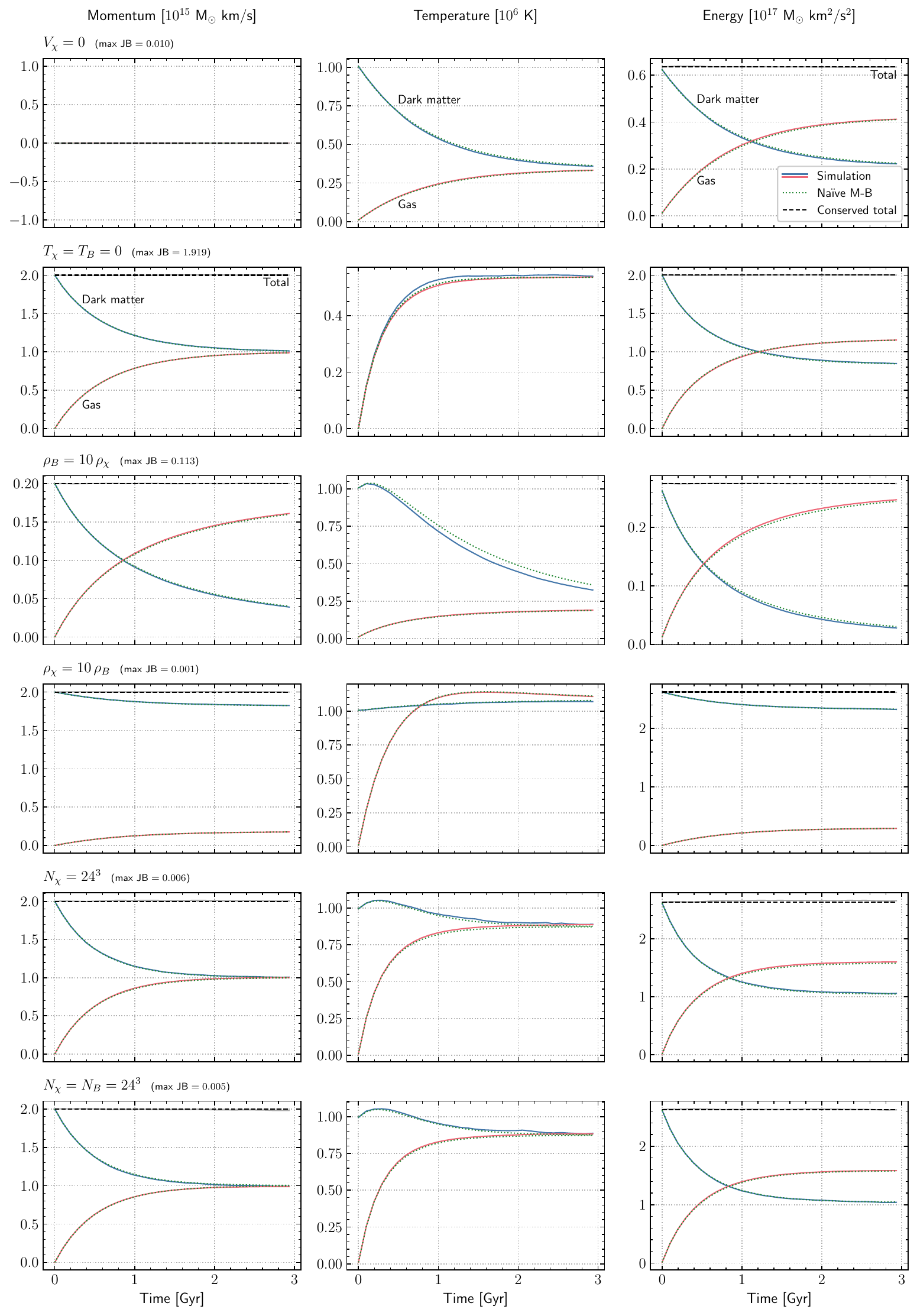}
    \caption{As in \Fig\ref{fig:tests-fiducial}, except varying the initial bulk velocity (top), temperatures (second row), densities (third and fourth), and resolutions (fifth and sixth).}
    \label{fig:tests-standard}
\end{figure}

\begin{figure}
    \centering
    \includegraphics[width=0.83\linewidth]{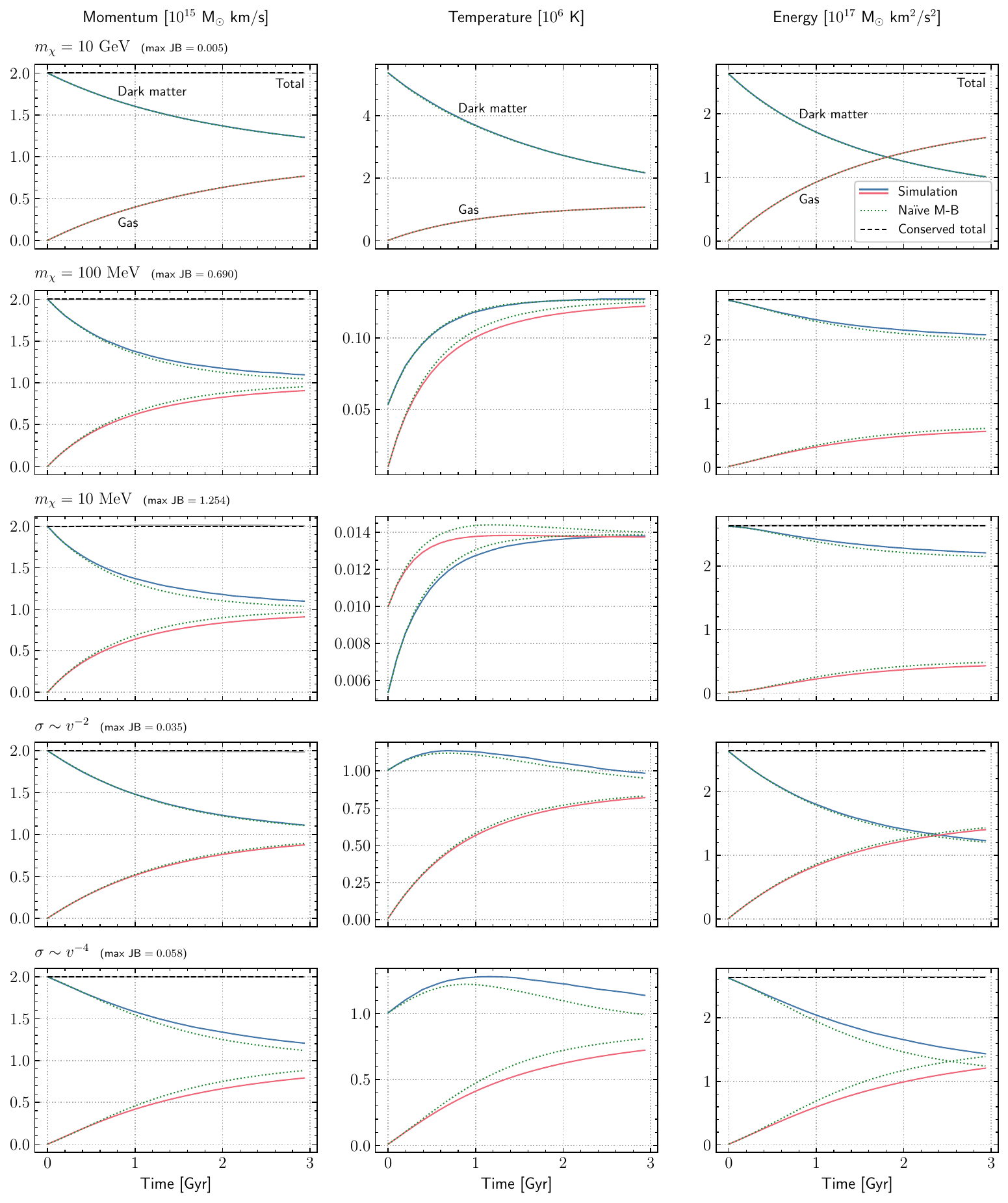}
    \caption{As in \Fig\ref{fig:tests-fiducial}, except varying the dark matter mass (first three rows) and velocity-dependence of the cross section (bottom two rows).}
    \label{fig:tests-different}
\end{figure}

The simulation results and na\"ive Maxwell-Boltzmann predictions are summarized in \Figs\ref{fig:tests-fiducial} through \ref{fig:tests-different}, which show the evolution of the momenta (left panel), temperatures (middle), and energies (right).
Each panel shows
    the simulation results in blue (dark matter), pink (gas), and grey (total) solid lines,
    the na\"ive M-B prediction in a dotted green line,
    and the initial value of the total momentum or energy as a black dashed line.
The maximum value of the Jarque-Bera statistic during the simulation is also reported for each test.
\Fig\ref{fig:tests-fiducial} shows the ``fiducial'' simulation, run with ten different random seeds in both the initial conditions generator and the seed inside GIZMO.
These help to give an idea of the scatter in the results due to the stochasticity of our method.
\Fig\ref{fig:tests-standard} shows the tests that vary the experimental setup (velocities, temperature, densities, resolutions), all using the same dark matter model: mass $m_\chi = 2 \, m_p$, velocity-independent cross section $\sigma = 10 \ {\rm mb}$.
\Fig\ref{fig:tests-different} shows the remaining tests, which vary the dark matter model but all use the same experimental setup (note that we change the initial dark matter temperature to keep the velocity dispersion fixed).

In all simulations except the reduced dark matter resolution ($N_\chi = 24^3$), we see excellent conservation of the global momentum and energy,
and the lack of perfect conservation in this one simulation is consistent with increased stochasticity because of the reduced resolution.
All simulations show sensible behavior, with the dark matter and baryons equilibrating in both momentum and temperature.
In all cases where the dark matter remains relatively Maxwell-Boltzmann (empirically, when $\text{max JB} \lesssim 0.01$), we see excellent agreement between the simulation and the M-B prediction.
For tests that attain larger JB values, the M-B prediction deviates from the simulation, usually predicting faster energy and momentum transfer than is observed in the simulation.
We stress that this is a \textit{failure of the na\"ive Maxwell-Boltzmann prediction} and not of the simulation: for these test setups, the dark matter really does deviate from a Maxwell-Boltzmann distribution.
Importantly, this suggests that assuming a Maxwell-Boltzmann distribution can be misleading for situations that arise in the evolution of galaxies.
This has been similarly pointed out for the effects of dark matter--baryon interactions on linear cosmology observables \citep{Ali-Haimoud2019}.

\begin{figure}
    \centering
    \includegraphics[width=0.48\linewidth]{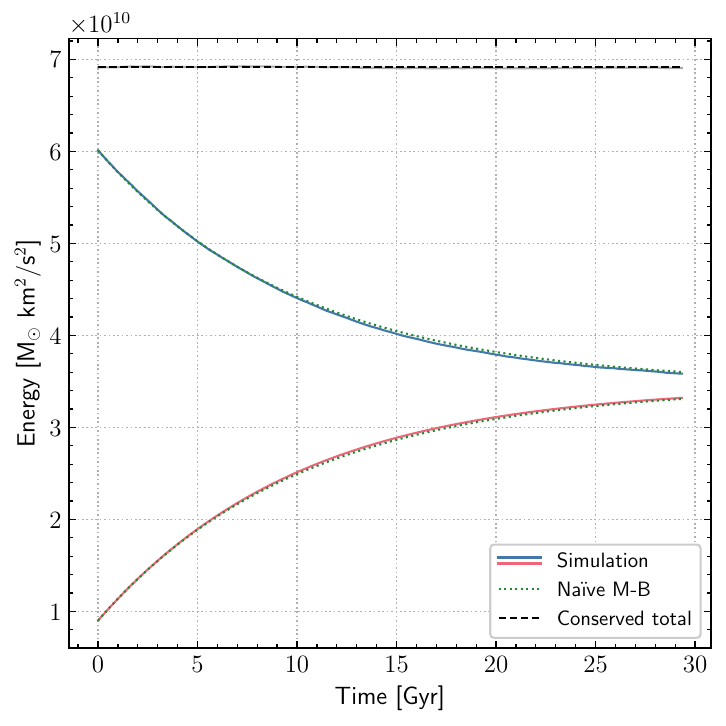}
    \includegraphics[width=0.48\linewidth]{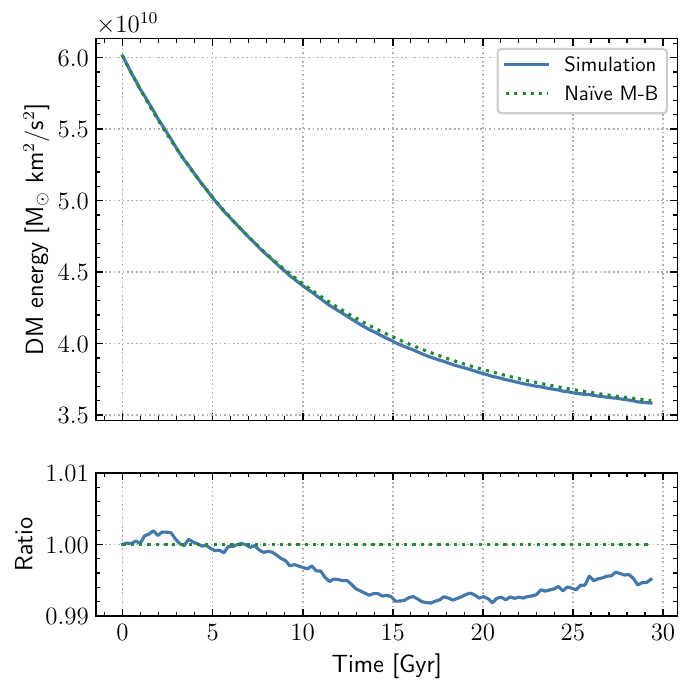} \\
    \includegraphics[width=0.48\linewidth]{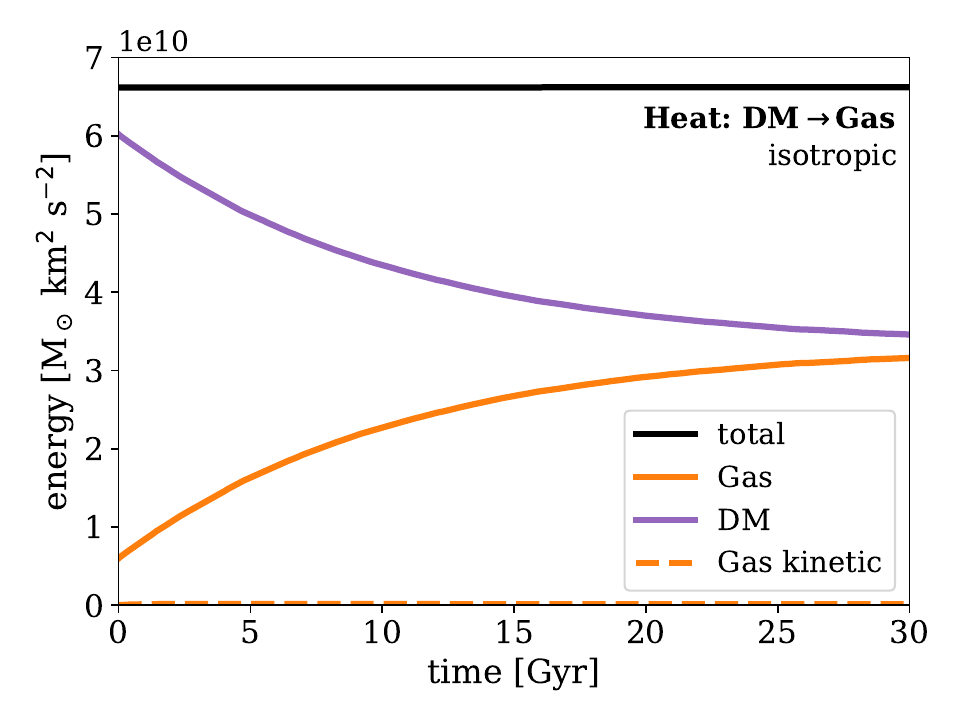}
    \includegraphics[width=0.48\linewidth]{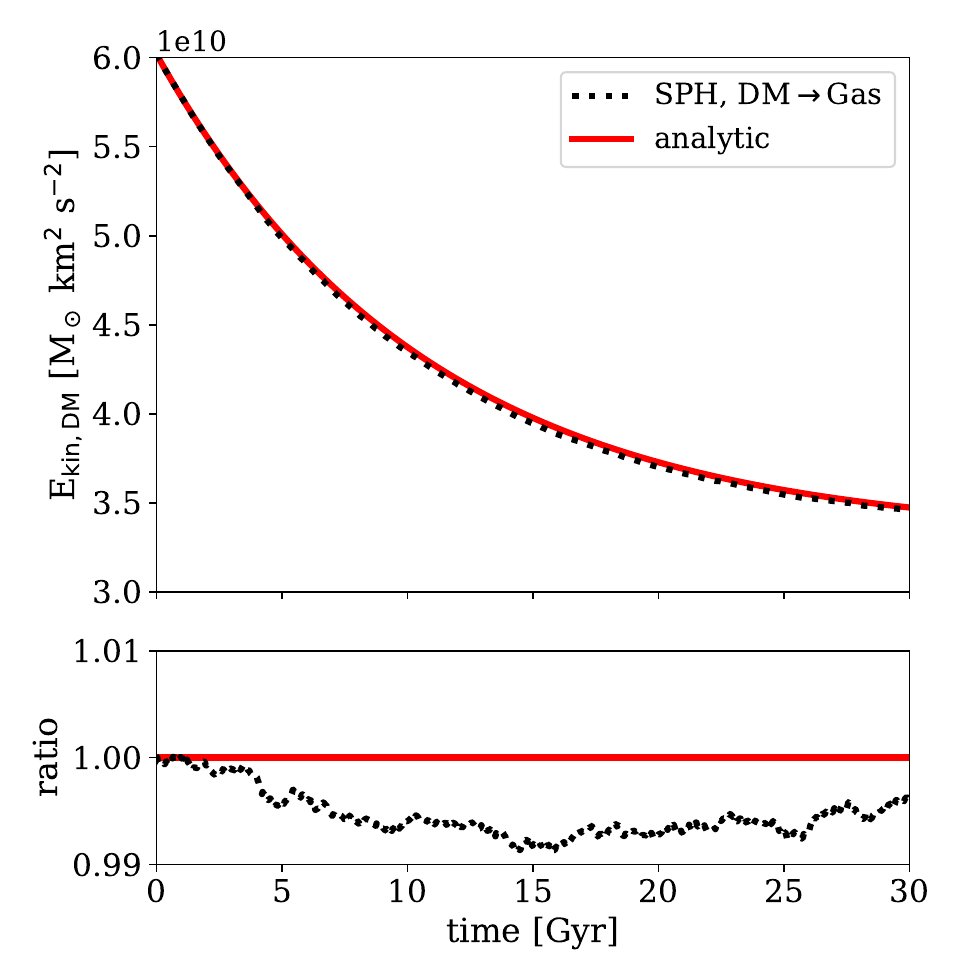}
    \caption{
    (Top row) Energetics of the ``F25 \S 3.4'' test problem, which re-creates the test detailed in Section 3.4 of \citet{Fischer+2025}.
    These plots are nearly indistinguishable from their Figures 12 and 13 (bottom row).
    }
    \label{fig:test_fischer}
\end{figure}

Serendipitously, \citet{Fischer+2025} used an almost identical test setup, and included one test simulation using a constant cross section with isotropic scatterings.
It was straightforward to re-create that simulation with our pipeline.
The parameters used are listed in \Tab\ref{tab:test_params} in the final row, titled ``F25 \S 3.4'':
briefly, compared to our fiducial test, the \Fischer{} test features proton-mass dark matter, a much larger cross section, smaller densities and temperatures, lower resolution, and a longer simulation time.
\Fig\ref{fig:test_fischer} shows the results.
Because the bulk velocities in this problem are zero, the global momentum plot contains no information and the temperature and energy plots are entirely redundant, so we show only the energies.
We also include a zoomed-in plot of the dark matter energy and its ratio versus the na\"ive Maxwell-Boltzmann prediction.
The behavior is indistinguishable from that reported in Figures 12 and 13 of \citet{Fischer+2025}, including the deviation of the simulated result from the na\"ive Maxwell-Boltzmann-estimated dark matter energy.

\end{document}